\RequirePackage{fix-cm}
\documentclass[onecolumn]{svjour3}          % twocolumn
\smartqed  % flush right qed marks, e.g. at end of proof
\usepackage{graphicx}
 \usepackage{mathptmx}      % use Times fonts if available on your TeX system
 \usepackage{amsmath}
 \usepackage{mathtools}
 \usepackage{lipsum}
 \usepackage{amssymb}
 \usepackage[english]{babel}
\usepackage[square,numbers]{natbib}
 \journalname{}
\begin{document}

\title{A discrete cohesive zone model for beam element: Application to adhesively bonded laminates and sandwich panels%\thanks{Grants or other notes
%about the article that should go on the front page should be
%placed here. General acknowledgments should be placed at the end of the article.}
}
%\subtitle{Do you have a subtitle?\\ If so, write it here}

%\titlerunning{Short form of title}        % if too long for running head

\author{Himanshu         \and
        Ananth Ramaswamy %etc.
}

%\authorrunning{Short form of author list} % if too long for running head

\institute{Himanshu \and Ananth Ramaswamy \at
              Department of Civil Engineering\\
              Indian Institute of Science Bangalore, Karnataka, India\\
              \email{himanshu1@iisc.ac.in}  \\
              \email{ananth@iisc.ac.in}
%             \emph{Present address:} of F. Author  %  if needed
}

\date{Received: date / Accepted: date}
% The correct dates will be entered by the editor

\maketitle

\begin{abstract}
A new discrete cohesive zone model (DCZM) is presented for modeling the interface behavior of adhesive-bonded thin laminates and sandwich panels. The proposed model treats the interface as a spring element and the adherent as a beam element. The use of the preceding assumptions facilitates the simplification of the computational framework reducing the problem from a 2D to 1D, thereby relaxing the requirements of maintaining the aspect ratio of elements in the finite element mesh. For thin laminates, the constitutive relation of the adhesive is represented by a bi-linear traction-separation law, whereas for sandwich panels, an exponential law is employed to model the adhesive behavior. In order to validate the proposed model for thin laminates, simulations of three established fracture tests: the double cantilever beam (DCB), an end notch flexure (ENF) beam, and mixed-mode beam (MMB) have been undertaken. Additionally, for the sandwich panel, two experiments documented in the literature have been simulated for assessing the efficacy of the modelling. One experiment has a mode-I interface failure and the other one a mode-II interface failure between the core and skin. It has been observed that the model is not sensitive to either the element size or the load step size. The results have been compared with reported (benchmark) numerical, analytical, and experimental findings. The proposed methodology for thin laminates offers a significant reduction in the computational effort (reduced number of unknown degrees of freedom compared to existing methods) with no compromise on the accuracy of the predictions. Specifically, it reduces the unknown degrees of freedom by more than 25\% compared to the corresponding mesh used in existing continuum cohesive zone model (CCZM) approaches. The Newton-Raphson solver can achieve quick convergence and no line search feature is required. The proposed algorithm is easy to implement on any computational platform.
% \PACS{PACS code1 \and PACS code2 \and more}
%\subclass{MSC code1 \and MSC code2 \and more}
\end{abstract}

\section{Introduction}
\label{intro}
Composite constructions have acquired universal recognition in the automobile, aerospace, marine, and sustainable energy industries. Composites provide many advantages over bulk metal, such as a high strength and stiffness-to-weight ratio, low cost, thermal insulation, and durability. Laminate and sandwich panels are prevalent composite structures that are extensively used in engineering applications to fulfill essential requirements owing to its enhanced bending stiffness, in-plane stiffness, and strength. Fracture in composite laminates and sandwich structures is one of the interesting areas that has been explored by many researchers over the past few decades. In composite laminates, common types of failure include the breakage of the matrix and fibers, debonding of fibers from the matrix, and delamination between lamina ~\cite{zimmermann2020review}, whereas for sandwich panels, common types of failure include wrinkling of the skin, shear failure of the core, failure or yielding of the skin, and delamination of the skin from the core~\cite{r39}. Delamination of both, lamina in the case of laminate and skin from the core in the case of a sandwich panel can be seen as a fracture mechanics problem.\\
\\ Linear Elastic Fracture Mechanics (LEFM), as introduced by Griffith~\cite{griffith1921vi}, is a promising theory for analyzing the fracture mechanics of structures. To enhance the generality of the theory, the principles of LEFM are integrated with the Finite Element Method (FEM) in situations where analytical solutions are difficult~\cite{banks1991application,pang1993linear}. The Extended Finite Element Method (XFEM) emerges as a notable numerical technique, enriching the finite element space with discontinuous functions to accurately capture the local mechanical behavior around the crack tip~\cite{moes1999finite}. Additionally, the application of XFEM allows for the investigation of the normal direction of the fracture surface under various failure modes in laminate composites~\cite{jia2021failure}. Furthermore, XFEM is employed to model the delamination phenomenon between laminae~\cite{zhao2016xfem},
~\cite{abdullah2017transversal}, ~\cite{yazdani2016xfem}. However, XFEM is constrained to situations where a pre-existing crack in the body is present, and it does not simulate the crack nucleation phenomenon, primarily because the LEFM theory does not deal with crack initiation. Additionally, due to computational costs, XFEM elements are typically utilized only in regions where the crack path is known in advance. Cohesive zone model(CZM) is one of the efficient model that overcome all the above mentioned limitation.\\
\\The foundation of CZM can be traced to the Barenblatt-Dugdale (BD) model that was initially developed for steel \cite{r3}. A process zone ahead of the crack tip is considered in the BD model, where stress is limited by material yield strength, resulting in a plastic strip of constant stress ahead of the intended crack path ~\cite{r4,r5,r6,r7}. Later the idea of the BD model has been extended towards concrete, to model the formation and growth of cracks. Instead of defining a yield strength in the process zone, a cohesive degradation law that defines material strength degradation as the strain ahead of crack tip advances is proposed~\cite{r8,r9,r10}. The cohesive degradation law, commonly known as the Traction Separation Law (TSL) is a material property, and its shape is generally dependent on whether the material is ductile or brittle~\cite{r19}. The CZM is strongly dependent on the length scale parameter($L_{c}$) that is a function of elastic modulus, Fracture toughness, and cohesive strength. Hence, for accurate results the mesh size is desired to be smaller than the $L_c$. However, in the present work, $L_c$ doesn't play a role on mesh size. \\
\\The CZM model, known for its ease of implementation, finds applications in simulating various mechanical systems. In laminated composites, the delamination between two laminae is modeled using a zero thickness eight-noded isoparametric plain-strain continuum element at the interface~\cite{r20}. However, this approach incurs significant computational costs, requiring a path-following technique like the arc length method for convergence~\cite{r16}. Depending on the system, the stiffness matrix may contain off-diagonal terms. Maintaining aspect ratio becomes challenging as the separation distance leading to softening behavior is considerably smaller than the element size of the adherent. This approach is commonly known as the Continuum Cohesive Zone Model (CCZM). Commercial software such as ABAQUS~\cite{r30} and ANSYS provides the capability to implement CCZM when the adherent thickness is minimal compared to its other dimensions, and achieving convergence is challenging due to a large number of degrees of freedom~\cite{r20}. Moreover, the model has been advanced to address various issues. For instance, the interface between fibers and matrix in SiC/Ti composites under off-axis loading conditions is successfully modeled~\cite{r35}. A similar approach is then applied to simulate the peeling test in steel-polymer composites~\cite{r36}. Fiber bridging plays a crucial role in the delamination process of fiber composites. To incorporate the impact of fiber bridging, a two-bilinear cohesive zone model (CZM) is utilized, where one segment represents matrix fracture, and another segment delineates the fiber bridging phenomenon ~\cite{r37}. Additionally, CZM has proven effective in laminates for simulating inter-laminar damage mechanisms ~\cite{r38}. Similar kind of delamination failures can be easily seen in sandwich panels. \\ 
\\A sandwich panel is composed of three components: the top skin, core, and bottom skin. These components are securely joined together using an adhesive bond. One of the primary modes of failure in a sandwich panel is interface failure, also referred to as delamination, which occurs between the core and skin layers~\cite{r39}.The pioneering work conducted by ~\cite{r40} involved the experimental investigation of mode-I debonding failure between the skin and core of a sandwich panel. The study revealed that the crack propagation from the interface to the core can occur, and this phenomenon is influenced by the core's density. ~\cite{r40} also reported a critical load at which the crack propagates steadily within the interface. During the experiment, the specimen was unloaded and reloaded after each finite crack growth increment. The debonding experiment of a Cracked Sandwich Beam (CSB) specimen subjected to three-point bending was conducted by ~\cite{r41}. The CSB specimen featured a pre-existing delamination crack located at the interface between the top skin and core near the support of the beam. Under applied loading, the crack predominantly propagated due to shear, resulting in a mode-II type failure. \\
\\~\cite{r43} introduced an advanced higher order beam theory that incorporates the compressibility of the soft core in the transverse direction. The developed theory considers the transverse displacement of the core as second order with respect to the transverse coordinate, while the in-plane displacement of the core is modeled as third order. The skin, on the other hand, is represented using the Timoshenko beam theory. This theory, known as "extended high-order sandwich panel theory" (EHSAPT), was successfully applied by ~\cite{r43}, demonstrating its capability to provide results in good agreement with existing elasticity findings reported in the literature. They assume a perfect bond between skin and core.~\cite{r44} proposed CCZM model to simulate the interface failure between the skin and core of a sandwich panel. While their model primarily focuses on mode-I loading conditions, it is important to note that other CCZM-based delamination models are also available in the literature ~\cite{r45}, ~\cite{r46}, ~\cite{r47}. However, CCZM is computationally expensive and some theories are specific to mode-I kind of loading condition. ~\cite{r43} developed a finite element based EHSAPT but the interfaces are assumed to be perfect. ~\cite{r49} has presented a nonlinear model for addressing the problem of interface crack propagation between the core and skin. However, their model involves a set of 26 coupled ordinary differential equations (ODEs), making it challenging to solve. To tackle this complexity, they employed finite difference methods to numerically solve the coupled equations. \\
\\As an alternate approach to overcome mesh dependency and convergence issue of CCZM, the discrete cohesive zone model(DCZM), has been developed~\cite{r17,r24,r25}. Instead of a zero-thickness continuum element, a zero-thickness rod type (or zero-thickness spring) has been used to model the interface. The properties of non-linear springs were developed based on the shape of the TSL and the mesh size of the adherent. The main idea is to concentrate the continuum properties of the interface on point-wise properties. DCZM is independent of mesh size and not sensitive to the loading increment~\cite{r24,r25}. The laminate thickness is usually significantly less than the other length, so the CCZM methodology is computationally expensive. These kinds of laminate composites are also bending dominant so modeling them as a continuum 2d element is not ideal. A de-lamination model for shell elements has been developed to overcome this issue. It consists of a cohesive zone model for degrading adhesive forces and an adhesive penalty contact formulation for initially connecting shells~\cite{r28}. Standard problems like DCB, ENF, and MMB specimens were solved using this technique, significantly reducing the computational cost.\\
\\In the present work, the proposed DCZM has two major differences compared to the formulation discussed in ~\cite{r24}. First, in~\cite{r24} and~\cite{r25} the adherent is modeled as a plain stress or strain element, while in the present  formulation, the adherent is modeled as a beam element. Second, in the present model, the stiffness of springs is independent of the interface's thickness. In~\cite{r24} the stiffness of springs is related to the thickness of the interface, which is subjective to choose, as the basic assumption in that study states that the thickness of the interface is zero.\\
\\For the thin laminates, the present methodology is implemented for standard mode-I, mode-II, and mix-mode problems. Comparison is made with previous work and existing analytical solutions. A bi-linear TSL is used. The proposed model is also scale-able with respect to the mesh size, as shown subsequently. As a result, for thin laminates the proposed model can greatly reduce the number of unknown degrees of freedom.\\
\\In the specific context of a sandwich beam, we have developed a novel finite element model that is based on the Extended High-Order Sandwich Panel Theory (EHSAPT). The proposed model effectively captures the interface between the core and the skin by employing a traction-separation law (TSL) proposed by ~\cite{r48}. Our proposed model is capable of simulating two distinct loading conditions: mode-I and mode-II. The primary objective of these simulations is to gain valuable insights into the behavior and performance of the sandwich beam under both loading scenarios. The methodology is straightforward to implement and can be programmed using any open-source language. For this particular study, MATLAB has been utilized.\\
\section{Methodology}
\subsection{Composite laminates}
\begin{figure}[h!]
    \centering
    \includegraphics[scale = 0.40]{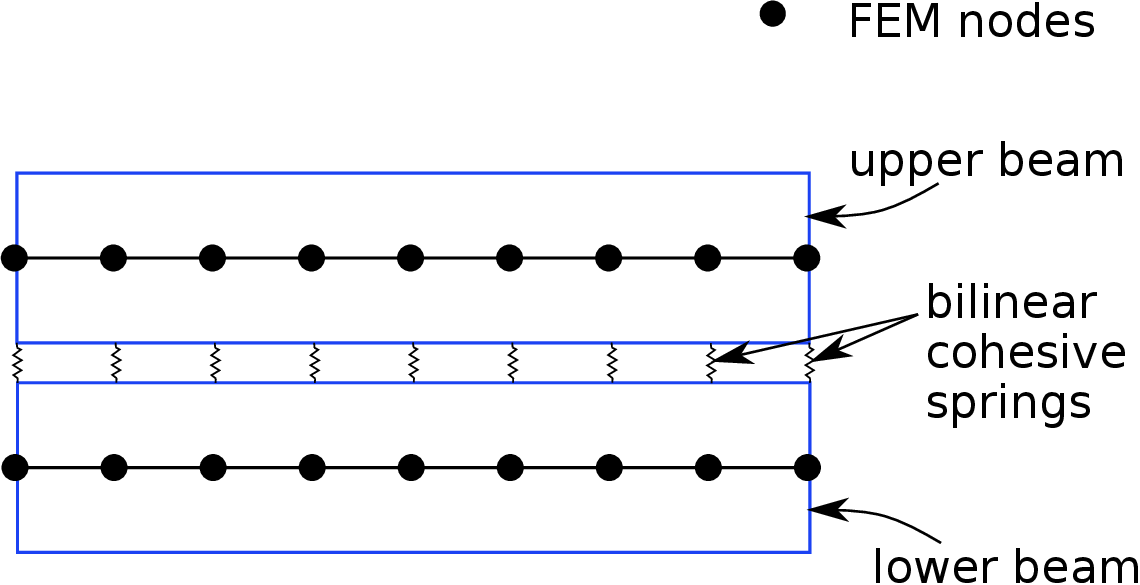}
    \caption{Meshing of assembly in un-deformed configuration}
    \label{fig:f1}
\end{figure}

A bi-linear spring is used in both longitude and transverse directions to model the interface(adhesive) between the two laminae. The adherent laminae are modeled as an 'Euler-Bernoulli' beam element with the degrees of freedom\( (U_1, U_2, \theta)\). Figure \ref{fig:f1} depicts the meshing of the assembly in the material configuration.\\ 

\begin{figure}[h!]
    \centering
    \includegraphics[scale = 0.4]{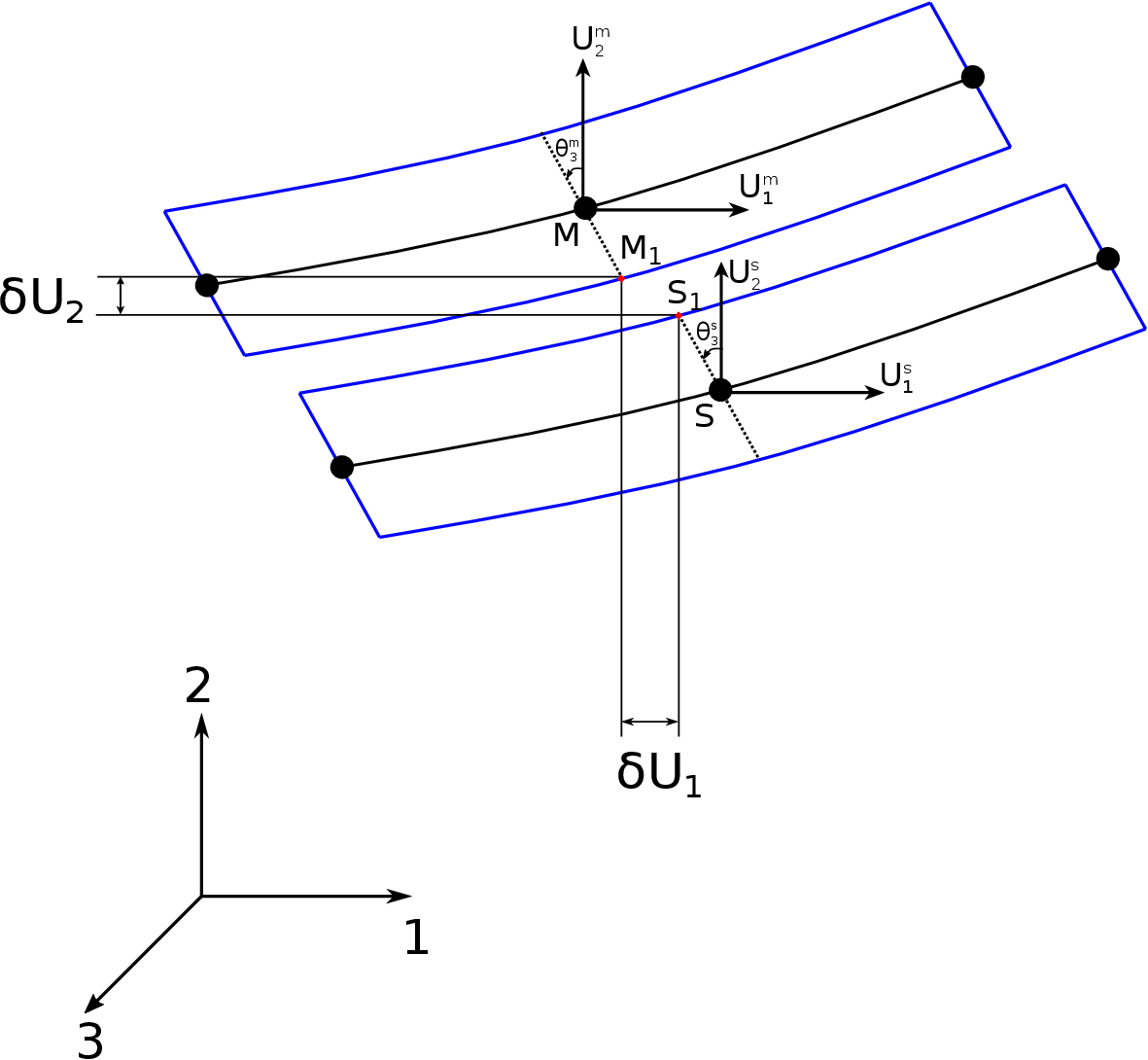}
    \caption{deformed configuration of beam assembly}
    \label{fig:f2}
\end{figure}

\begin{figure}[h!]
    \centering
    \includegraphics[scale = 0.20]{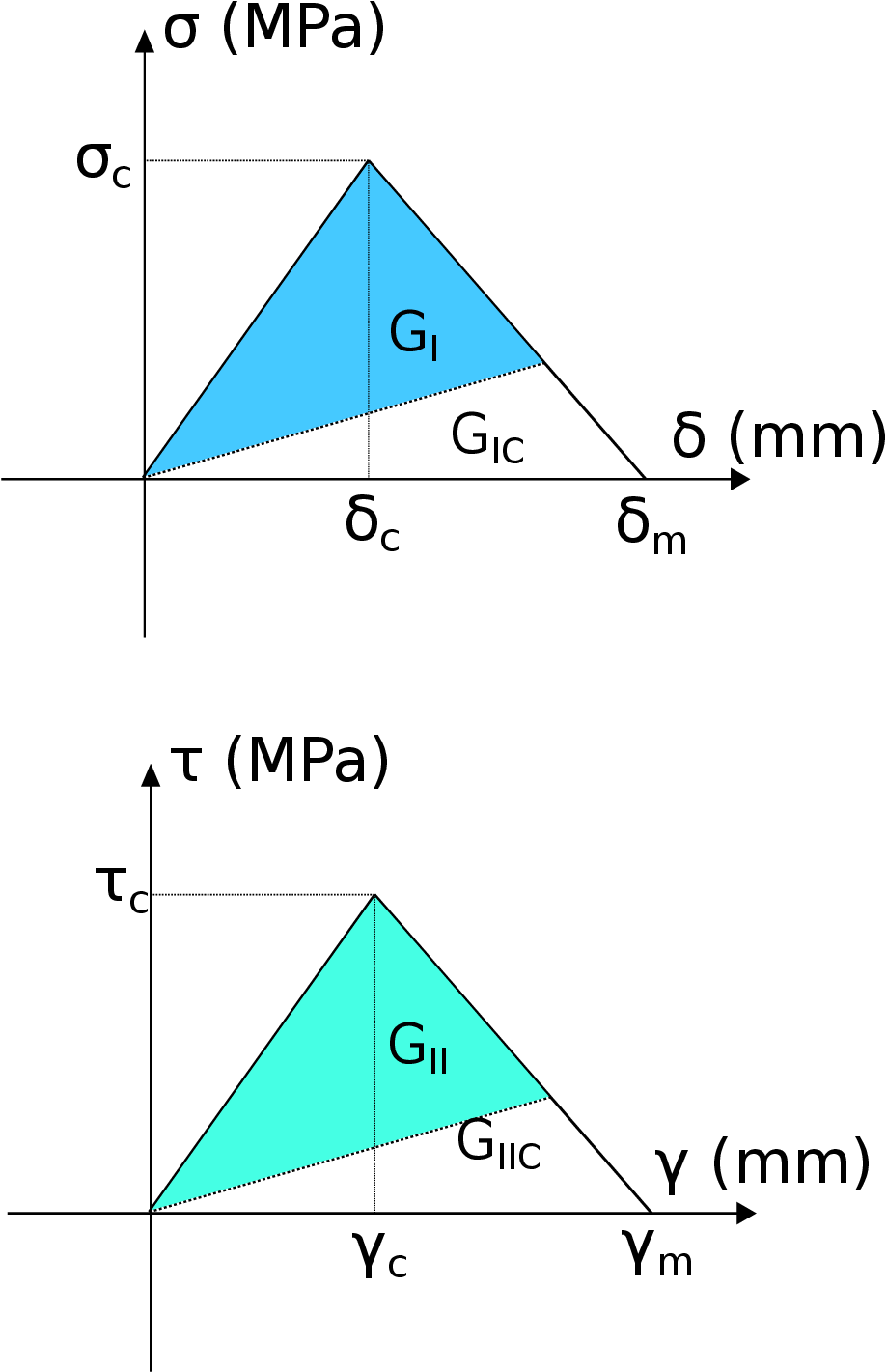}
    \caption{Traction separation law for Mode-I and Mode-II}
    \label{fig:f3}
\end{figure}

As shown in figure~\ref{fig:f1} the bottom surface of the upper beam is bonded with the top surface of the lower beam. As shown in figure~\ref{fig:f2} the contact points of the springs are labeled $ M_1$ and $S_1 $ and the relative displacement of these two points in the longitudinal and transverse directions determines the force in these springs.\\
$U^m_1$, $U^m_2$, $U^s_1$, and $U^s_2$ are the displacement component of node M and S along the axis 1 and 2 as shown in figure~\ref{fig:f2}. While $\theta^m_3$ and $\theta^s_3$ are the rotations of corresponding normal of node M and S about axis 3. The relative displacement component along 1 and 2 axis between $M_1$ and $S_1$ is given by
\begin{subequations}
\begin{align}\label{1}
%\begin{split}
\delta U_1 &= U^m_1 - U^s_1 - t_m\theta^m_3 - t_s\theta^s_3\\
\delta U_2 &= U^m_2 - U^s_2
%\end{split}
\end{align}
\end{subequations}
 where $2t_m$ and $2t_s$ are the thickness of master and slave beams. Forces developed in springs also depend upon the material properties of the interface. As shown in figure~\ref{fig:f3} triangular traction separation law defines the material properties of the interface. The energy required to create a new surface is given as
 \begin{subequations}
\begin{align}\label{2}
 G_{Ic} &= \frac{1}{2}\sigma_c\delta_c\\
 G_{IIc} &=\frac{1}{2}\tau_c\gamma_c
\end{align}
\end{subequations}
\\$G_{Ic}$ and $G_{IIc}$ are material properties measured by performing experiments~\cite{r11}. $G_{Ic}$ is typically measured with a D.C.B specimen, $G_{IIc}$ is measured with an E.N.F test, and mixed-mode fracture studies are performed with the M.M.B device. Inferring from Newton's third law, the forces at $M_1$ and $S_1$ would be equal and opposite. The maximum stress that the interface can take in mode I and II is a material parameter given as $\sigma_c$ and $\tau_c$. So the maximum force in each spring in terms of $\sigma_c$,  $\tau_c$ and mesh parameters is given as
 \begin{subequations}
\begin{align}\label{3}
 F_{Ic} &= \sigma_c\Delta a B\\
 F_{IIc} &=\tau_c\Delta a B
\end{align}
\end{subequations}
\\ B: the thickness of beam along direction 3
\\ $\Delta a$ : mesh size of beam element\\ 
\\upon using (3) the initial linear stiffness of springs can be computed in terms of $\sigma_c$ and $\gamma_c$. 
\begin{subequations}
\begin{align}\label{4}
 K_{I} &= \frac{\sigma_c\Delta a B}{\delta_c}\\
 K_{II} &=\frac{\tau_c\Delta a B}{\gamma_c}
\end{align}
\end{subequations}
\\On using equations (1) and (4) forces can be written in terms of unknown d.o.fs\( (U_1, U_2, \theta)\). These forces are equal and opposite at $M_1$ and $S_1$.\\
\begin{subequations}
\begin{align}\label{5}
F_1^m &= K_{II}\delta U_1\\
F_2^m &= K_{I}\delta U_2
\end{align}
\end{subequations}
The forces from  $M_1$ and $S_1$ were transferred to the relevant nodes on the neutral axis of beams at $M$ and $S$ as shown in figure 2. The moment generated at the nodes as a result of force transmission is given by
\begin{subequations}
\begin{align}\label{6}
M_3^m &= -K_{II}\delta U_1 t_m\\
M_3^s &= -K_{II}\delta U_1 t_s
\end{align}
\end{subequations}
 Finite element formulation has been implemented and the connection matrix that defines the interface between the corresponding nodes of two beams is given by
\begin{equation}\label{7}
%$
%{\footnotesize
%\left[\begin{array}{cccccc}
\begin{bmatrix}
K_{II} & 0 & -K_{II} t_s & -K_{II} & 0 & -K_{II} t_m\\
0 & K_{I} & 0 & 0 & -K_{I} & 0\\
-K_{II} t_s & 0 & K_{II} t_s^2 & K_{II} t_s & 0 & K_{II} t_m t_s\\
-K_{II} & 0 & K_{II} t_s & K_{II} & 0 & K_{II} t_m\\
0 & -K_{I} & 0 & 0 & K_{I} & 0\\
-K_{II} t_m & 0 & -K_{II} t_s t_m & -K_{II} t_m & 0 & -K_{II} t_m^2\\
\end{bmatrix}
%\end{array}\right]
%\left\{\begin{array}{c}
\begin{Bmatrix}
U_1^s\\
U_2^s\\
\theta_3^s\\
U_1^m\\
U_2^m\\
\theta_3^m\\
\end{Bmatrix}
%\end{array}\right\}
=
%\left\{\begin{array}{c}
\begin{Bmatrix}
F_1^s\\
F_2^s\\
M_3^s\\
F_1^m\\
F_2^m\\
M_3^m\\
\end{Bmatrix}
%\end{array}\right\}
%}
\end{equation}
\\The analysis is carried out using the Newton-Raphson iteration approach. From the previous converged time step the relative displacement between $M_1$ and $S_1$ ($\delta U_1$ and $\delta U_2$) is calculated. Let's consider mode-I case \\
\\If  $\delta U_2 < \delta_c$(undamaged state) the stiffness of spring, damage and energy release rate for the next time step is given as\\
\begin{equation}\label{8}
K_{I} = \frac{\sigma_c\Delta a B}{\delta_c} 
\end{equation}
\begin{equation}\label{9}
d_{I} = 0 
\end{equation}
\begin{equation}\label{10}
G_{I} = 0
\end{equation}
if $\delta_c < \delta U_2 < \delta_m$(partially damaged)
\begin{equation}\label{11}
K_{I} = -\frac{\sigma_{c} B \Delta a}{(\delta_m - \delta_c)} 
\end{equation}
\begin{equation}\label{12}
d_{I} = 1 - \frac{\delta_c(\delta_m - \delta U_2)}{\delta U _2 (\delta_m - \delta_c)}
\end{equation}
\begin{equation}\label{13}
G_{I} = \frac{\delta_c \delta U_2 K_{I} d_{I}}{2}
\end{equation}
if $\delta U_2 > \delta_m$(fully damaged)
\begin{equation}\label{14}
K_{I} = 0
\end{equation}
\begin{equation}\label{15}
d_{I} = 1 
\end{equation}
\begin{equation}\label{16}
G_{I} = G_{Ic} 
\end{equation}
similar, equations have also been used for the Mode-II condition. For the Mixed Mode case, Griffith's criteria are used,
\begin{equation}\label{17}
\frac{G_{I}}{G_{Ic}} + \frac{G_{II}}{G_{IIc}} \geq 1 
\end{equation}
Once this criterion is achieved the stiffness $K_{I}$ and $K_{II}$ is set equal to zero, this implies the advancement of the crack. \\
\\The normal displacement gap($\delta U_2$) at the interface may become negative in the  E.N.F and M.M.B specimens. To avoid a non-physical phenomenon of overlapping of the upper and lower beam, a contact condition is used
 \begin{equation}\label{18}
 F_2^m = K_{c}\delta U_2H(-\delta U_2)
 \end{equation}
 where H: Heaviside function.\\
$k_{c}$ : a spring coefficient with larger value.\\
\subsection{Sandwich panel}
\begin{figure}[h!]
    \centering
    \includegraphics[scale = 0.40]{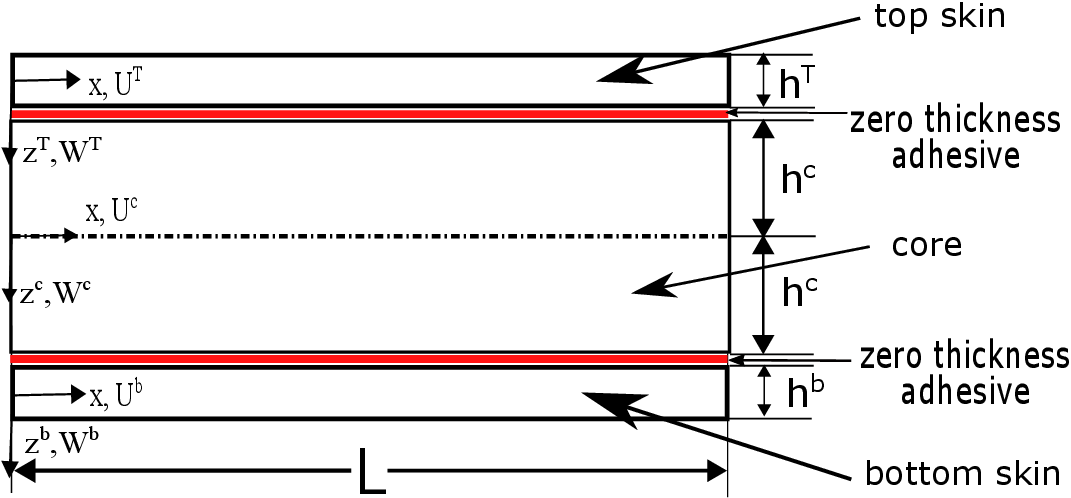}
    \caption{Structure of a sandwich panel}
    \label{fig:f4}
\end{figure}

\begin{figure}[h!]
    \centering
    \includegraphics[scale = 0.25]{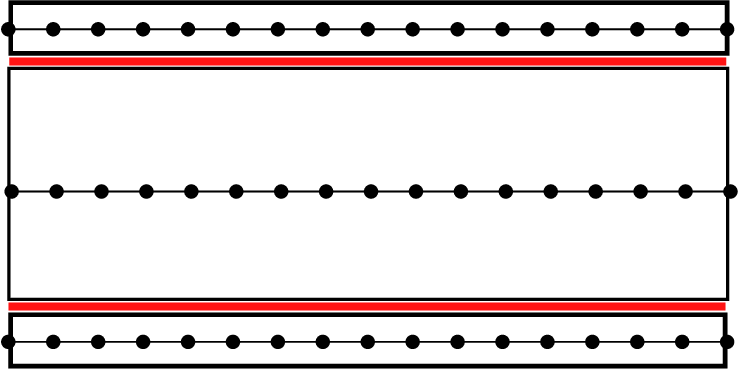}
    \caption{meshing of a sandwich panel}
    \label{fig:f5}
\end{figure}

\begin{figure}[h!]
    \centering
    \includegraphics[scale = 0.40]{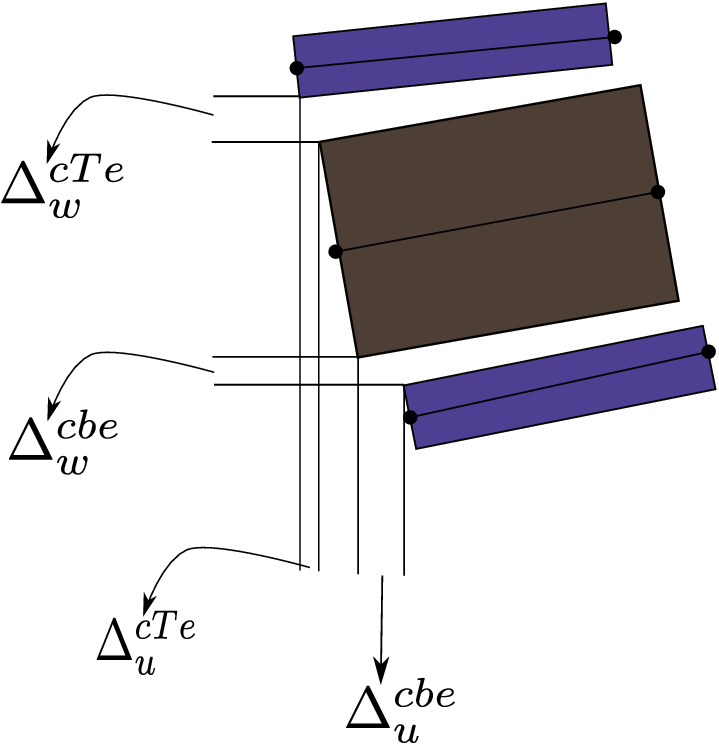}
    \caption{Opening of interfaces at the $e^{th}$ node }
    \label{fig:f6}
\end{figure}

\begin{figure}[h!]
    \centering
    \includegraphics[scale = 0.40]{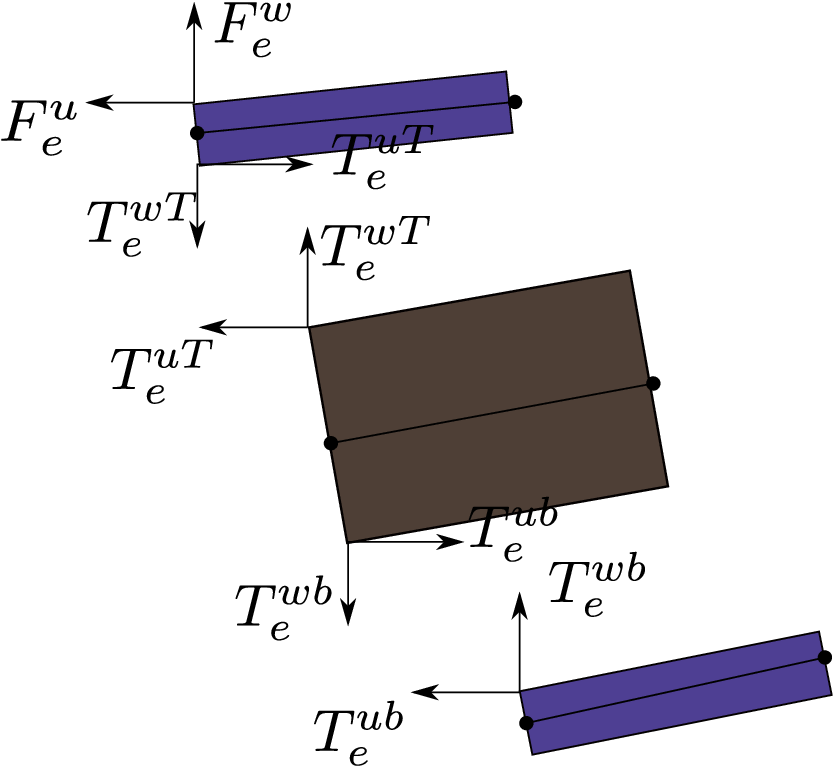}
    \caption{development of internal traction at the interfaces of $e^{th}$ node }
    \label{fig:f7}
\end{figure}

Consider a sandwich panel consisting of three layers: a top skin, a core, and a bottom skin. The core is bonded to the skins using a zero thickness adhesive, as illustrated in Figure ~\ref{fig:f4}. In order to model the interfacial debonding between the top skin and the core, a finite element formulation is proposed. For the purpose of this formulation, let's consider a sandwich beam where the top skin, core, and bottom skin are discretized using an equal number of two noded beam elements.
\\As shown in Figure ~\ref{fig:f5}, it is required that the nodes of the beam elements in the top skin, core, and bottom skin lie on the same vertical line. In this formulation, we assume that only the interface between the top skin and the core is susceptible to failure. The interface between the core and the bottom skin is also treated as an adhesive interface, but failure is not allowed in this specific formulation. It should be noted that the formulation can be extended to include a different number of layers, and each interface can be permitted to fail. However, in order to validate the formulation against existing experimental data in the literature, we restrict the failure analysis to the top interface only.\\
\\Consider the $e^{th}$ element of the top skin, core, and bottom skin, where each element consists of the $e^{th}$ and $(e+1)^{th}$ nodes. The displacement of the top and bottom skin in the transverse and axial directions can be expressed as:
\begin{equation}
    W^i(x,z) = w^i(x)
\end{equation}
\begin{equation}
    U^i(x,z) = u^i_o(x) - z^i\theta^i(x)
\end{equation}\\
where $i = T,b$ refers to the top and bottom skin, respectively.\\
The core is modeled using a higher order beam element, where the axial and transverse displacement are approximated using Taylor's series expansion. The assumed displacement expressions are as follows:
\begin{equation}
    U^c(x,z^c) = u_o^c(x) + z^cu_1^c(x) + (z^c)^2u_2^c(x) + (z^c)^3u_3^c(x)
\end{equation}
\begin{equation}
    W^c(x,z^c) = w_o^c(x) + z^cw_1^c(x) + (z^c)^2w_2^c(x) 
\end{equation}
\\Under static loading conditions, the assembly will deform, leading to the opening of interfaces between the top skin and core, as depicted in Figure ~\ref{fig:f6}.
\\The displacement jump over the $e^{th}$ node of the top skin and core can be defined as:
\begin{equation}
    \Delta_{u}^{cTe} = u_{o}^{ce} - h^{c}u_1^{ce} + (h^{c})^{2}u_{2}^{ce} - (h^{c})^3u_{3}^{ce} - u_e^{T} + \frac{h^{T}\theta_e^T}{2}
\end{equation}

\begin{equation}
    \Delta_{w}^{cTe} = w_{o}^{ce} - h^cw_1^{ce} + (h^c)^2w_2^{ce}-w_e^T
\end{equation}
\\where$\Delta_{u}^{cTe}$ represent the displacement jump at the $e^{th}$ node in the axial direction, and $\Delta_{w}^{cTe}$ denote the displacement jump in the transverse direction.\\
Similarly, displacement jumps for core and bottom skin at $e^{th}$ node can also be written as
\begin{equation}
\Delta_u^{cbe} = u_e^b + \frac{h^b\theta_e^b}{2} - u_{o}^{ce} - h^{c}u_1^{ce} - (h^{c})^{2}u_{2}^{ce} - (h^{c})^3u_{3}^{ce}
\end{equation}
\begin{equation}
    \Delta_w^{cbe} = w_e^{b} - w_{o}^{ce} - h^cw_1^{ce} - (h^c)^2w_2^{ce}
\end{equation}
The traction resulting from the displacement jump is modeled using the Needleman cohesive zone model ~\cite{r48}. As shown in Figure ~\ref{fig:f7}, internal tractions are developed at the interfaces.
\\Specifically, the traction developed at the interface between the top skin and core can be expressed as:

\begin{equation}
    T_e^{uT} = \frac{2\varphi^{cT}\Delta_u^{cTe}}{(\delta_o^{cT})^2}\left(1 + \frac{\Delta_w^{cTe}}{\delta_o^{cT}}\right)\exp\left(-\frac{\Delta_w^{cTe}}{\delta_o^{cT}} -\left(\frac{\Delta_w^{cTe}}{\delta_o^{cT}}\right)^2\right)\frac{b_wL_e}{2}
\end{equation}

\begin{equation}
    T_e^{wT} = \frac{\varphi^{cT}\Delta_w^{cTe}}{(\delta_o^{cT})^2}\exp\left(-\frac{\Delta_w^{cTe}}{\delta_o^{cT}} -\left(\frac{\Delta_w^{cTe}}{\delta_o^{cT}}\right)^2\right)\frac{b_wL_e}{2}
\end{equation}

where:

$T_e^{uT}$ represents the traction developed in the interface of the top skin and core along the x-direction at the $e^{th}$ node.

$T_e^{wT}$ represents the traction developed in the interface of the top skin and core along the z-direction at the $e^{th}$ node.

$\varphi^{cT}$ denotes the work of separation per unit area of the interface between the top skin and core.

$\delta_o^{cT}$ represents the characteristic length scale parameter of the interface between the top skin and core.

$b_w$ denotes the width of the beam.

$L_e$ represents the length of the beam element.

Since we do not allow damage to occur at the interface between the core and bottom skin, the traction developed at the $e^{th}$ node of that interface can be given as:

\begin{equation}
    T_e^{ub} = \frac{2\varphi^{cb}\Delta_u^{cbe}}{(\delta_o^{cb})^2}\frac{b_wL_e}{2}
\end{equation}

\begin{equation}
    T_e^{wb} = \frac{\varphi^{cb}\Delta_w^{cbe}}{(\delta_o^{cb})^2}\frac{b_wL_e}{2}
\end{equation}

where:

$T_e^{ub}$ represents the traction developed in the interface of the core and bottom skin along the x-direction at the $e^{th}$ node.

$T_e^{wb}$ represents the traction developed in the interface of the core and bottom skin along the z-direction at the $e^{th}$ node.

$\varphi^{cb}$ denotes the work of separation per unit area of the interface between the core and bottom skin.

$\delta_o^{cb}$ represents the characteristic length scale parameter of the interface between the core and bottom skin.
\\Considering that each element of the top skin, core, and bottom skin is stacked on top of each other, we can construct a comprising $e^{th}$ element of sandwich beam. This element provides information about the displacement jump between the $e^{th}$ and $(e+1)^{th}$ nodes when the top skin is subjected to a specific boundary condition. In the Appendix, we have derived the stiffness matrix for the higher order beam element used to model the core. Considering Figure \ref{fig:f6}, we can now formulate the residue force vector for the $e^{th}$ element of the upper skin. 
\begin{equation}
    \Delta R^{Te} = 
    \begin{bmatrix}
        [k_{e,e}^T]_{3*3} & [k_{e,(e+1)}^T]_{3*3}\\
        [k_{(e+1),e}^T]_{3*3} & [k_{(e+1),(e+1)}^T]_{3*3}
    \end{bmatrix}
   \begin{Bmatrix}
   \{\hat{u}^{T(e)}\}_{3*1}\\
   \{\hat{u}^{T(e+1)}\}_{3*1}      
   \end{Bmatrix} - 
   \begin{Bmatrix}
       \{T_e^T\}_{3*1}\\
       \{T_{e+1}^T\}_{3*1}\\
   \end{Bmatrix} - 
   \begin{Bmatrix}
       \{F_e^{ext}\}_{3*1}\\
       \{F_{e+1}^{ext}\}_{3*1}
   \end{Bmatrix}
\end{equation}
The displacement and force quantities associated with the $e^{th}$ node of the top skin and core are denoted by ${{\hat{u}^{T(e)}}_{3*1}}$ and ${T_e^T}_{3*1}$, respectively. These quantities represent the axial force, transverse force, and bending moment resulting from the interface jump between the top skin and core at the $e^{th}$ node of the top skin. They can be expressed as:
\begin{equation}
  \{T_e^T\}_{3*1} = \begin{Bmatrix}
      T_e^{uT}\quad T_e^{wT}\quad -\frac{h^T}{2}\left(T_e^{wT}\theta_e^T + T_e^{uT}\right)
  \end{Bmatrix}^t  
\end{equation}
similar expression holds for $(e+1)^{th}$ node.
By observing Figure 8, we can visualize the force balance of the core. The residue force vector for the $e^{th}$ element of the core assembly can be given as:
\begin{equation}
    \Delta R^{ce} = \begin{bmatrix}
        [k_{e,e}^c]_{7*7} & [k_{e,(e+1)}^c]_{7*7}\\
        [k_{(e+1),e}^c]_{7*7} & [k_{(e+1),(e+1)}^c]_{7*7}
                     \end{bmatrix}
                      \begin{Bmatrix}
                      \{\hat{u}^{c(e)}\}_{7*1}\\
                      \{\hat{u}^{c(e+1)}\}_{7*1}      
                       \end{Bmatrix} -
                       \begin{Bmatrix}
                        \{T_e^c\}_{7*1}\\
                        \{T_{e+1}^c\}_{7*1}\\
                        \end{Bmatrix}
\end{equation}
where $\{T_e^c\}_{7*1}$ is defined as: 
\begin{equation}
    \{T_e^c\}_{7*1} = \begin{Bmatrix}
        T_e^{ub}\\
        T_e^{ub}h^c\\
        T_e^{ub}(h^c)^2\\
        T_e^{ub}(h^c)^3\\
        T_e^{wb}\\
        T_e^{wb}h^c\\
        T_e^{wb}(h^c)^2
                     \end{Bmatrix}
                     -
                    \begin{Bmatrix}
        T_e^{uT}\\
        -T_e^{uT}h^c\\
        T_e^{uT}(h^c)^2\\
        -T_e^{uT}(h^c)^3\\
        T_e^{wT}\\
        -T_e^{wT}h^c\\
        T_e^{wT}(h^c)^2
                     \end{Bmatrix}
\end{equation}
similarly, the residue vector for the bottom skin of the $e^{th}$ element can be expressed as
\begin{equation}
    \Delta R^{be} = \begin{bmatrix}
        [k_{e,e}^b]_{3*3} & [k_{e,(e+1)}^b]_{3*3}\\
        [k_{(e+1),e}^b]_{3*3} & [k_{(e+1),(e+1)}^b]_{3*3}
    \end{bmatrix}
   \begin{Bmatrix}
   \{\hat{u}^{b(e)}\}_{3*1}\\
   \{\hat{u}^{b(e+1)}\}_{3*1}      
   \end{Bmatrix} - 
   \begin{Bmatrix}
       \{T_e^b\}_{3*1}\\
       \{T_{e+1}^b\}_{3*1}\\
   \end{Bmatrix}
\end{equation}
where,
\begin{equation}
  \{T_e^b\}_{3*1} = \begin{Bmatrix}
      -T_e^{ub}\quad -T_e^{wb}\quad -\frac{h^b}{2}\left(T_e^{wb}\theta_e^b + T_e^{ub}\right)
  \end{Bmatrix}^t  
\end{equation}
Taking the residues of top skin, core and bottom skin, rearranging them as shown below:
\begin{equation}
    \Delta R =\begin{bmatrix} K \end{bmatrix} 
    \begin{Bmatrix}
        U
    \end{Bmatrix} - \begin{Bmatrix}
        T
    \end{Bmatrix}
    - \begin{Bmatrix}
        F
    \end{Bmatrix}
\end{equation}
where 
\begin{equation}
    \begin{bmatrix}
        K
    \end{bmatrix} = 
%    {\tiny
    \begin{bmatrix}
        [k_{e,e}^T]_{3*3}&0&0&[k_{e,(e+1)}^T]_{3*3}&0&0\\
        0&[k_{e,e}^c]_{7*7}&0&0&[k_{e,(e+1)}^c]_{7*7}&0\\
        0&0&[k_{e,e}^b]_{3*3}&0&0&[k_{e,(e+1)}^b]_{3*3}\\
        [k_{(e+1),e}^T]_{3*3}&0&0&[k_{(e+1),(e+1)}^T]_{3*3}&0&0\\
        0&[k_{(e+1),e}^c]_{7*7}&0&0&[k_{(e+1),(e+1)}^c]_{7*7}&0\\
        0&0&[k_{(e+1),e}^b]_{3*3}&0&0&[k_{(e+1),(e+1)}^b]_{3*3}
    \end{bmatrix}
 %   }
\end{equation}
\begin{equation}
    \begin{Bmatrix}
        U
    \end{Bmatrix} = 
    \begin{Bmatrix}
           \{\hat{u}^{T(e)}\}_{3*1}\\
           \{\hat{u}^{c(e)}\}_{7*1}\\
           \{\hat{u}^{b(e)}\}_{3*1}\\
           \{\hat{u}^{T(e+1)}\}_{3*1}\\
           \{\hat{u}^{c(e+1)}\}_{7*1}\\
           \{\hat{u}^{b(e+1)}\}_{3*1}\\
    \end{Bmatrix}
\end{equation}
\begin{equation}
\begin{Bmatrix}
    T
\end{Bmatrix} =
\begin{Bmatrix}
           \{T_e^T\}_{3*1}\\
           \{T_e^c\}_{7*1}\\
           \{T_e^b\}_{3*1}\\
           \{T_{e+1}^T\}_{3*1}\\
           \{T_{e+1}^c\}_{7*1}\\
           \{T_{e+1}^b\}_{3*1}
\end{Bmatrix}
\end{equation}
using this residue vector tangent stiffness matrix for this sandwich element is given as
\begin{equation}
    K_{tang} = \frac{\partial \Delta R}{\partial \begin{Bmatrix}
        U
    \end{Bmatrix}}
\end{equation}
\\The residue and tangent stiffness matrix for each sandwich element can be calculated by following the formulation provided earlier. Once the residue and tangent stiffness matrices are obtained for each sandwich element, the full problem can be solved using the Newton-Raphson method.
\section{Results and discussion}
The proposed methodology was assessed by analyzing three common fracture test configurations: the D.C.B (Mode - I), E.N.F (Mode - II) and M.M.B (Mixed Mode). For all finite element modelling, Euler-Bernoulli beam elements are used. Figure 4 depicts the experimental setup for a D.C.B specimen. At point A, displacement is prescribed in the y-direction, and the corresponding reaction has been evaluated. The dimensions are length (2L) = $150mm$, Width(b) = $25mm$, initial crack length(a) = $35mm$, height(h) =$2.25mm$ young’s modulus(E) = $33.5N/mm^2$. Using experimental data the cohesive law parameters has been found by Leo~\cite{r31}. For adhesive a bi-linear traction separation law is used with properties $\sigma_c = 1.93N/mm^2$, $\delta_m = 0.6839mm$, and $G_{Ic} = 0.66N/mm$.\\
\begin{figure}[h!]
    \centering
    \includegraphics[scale = 0.40]{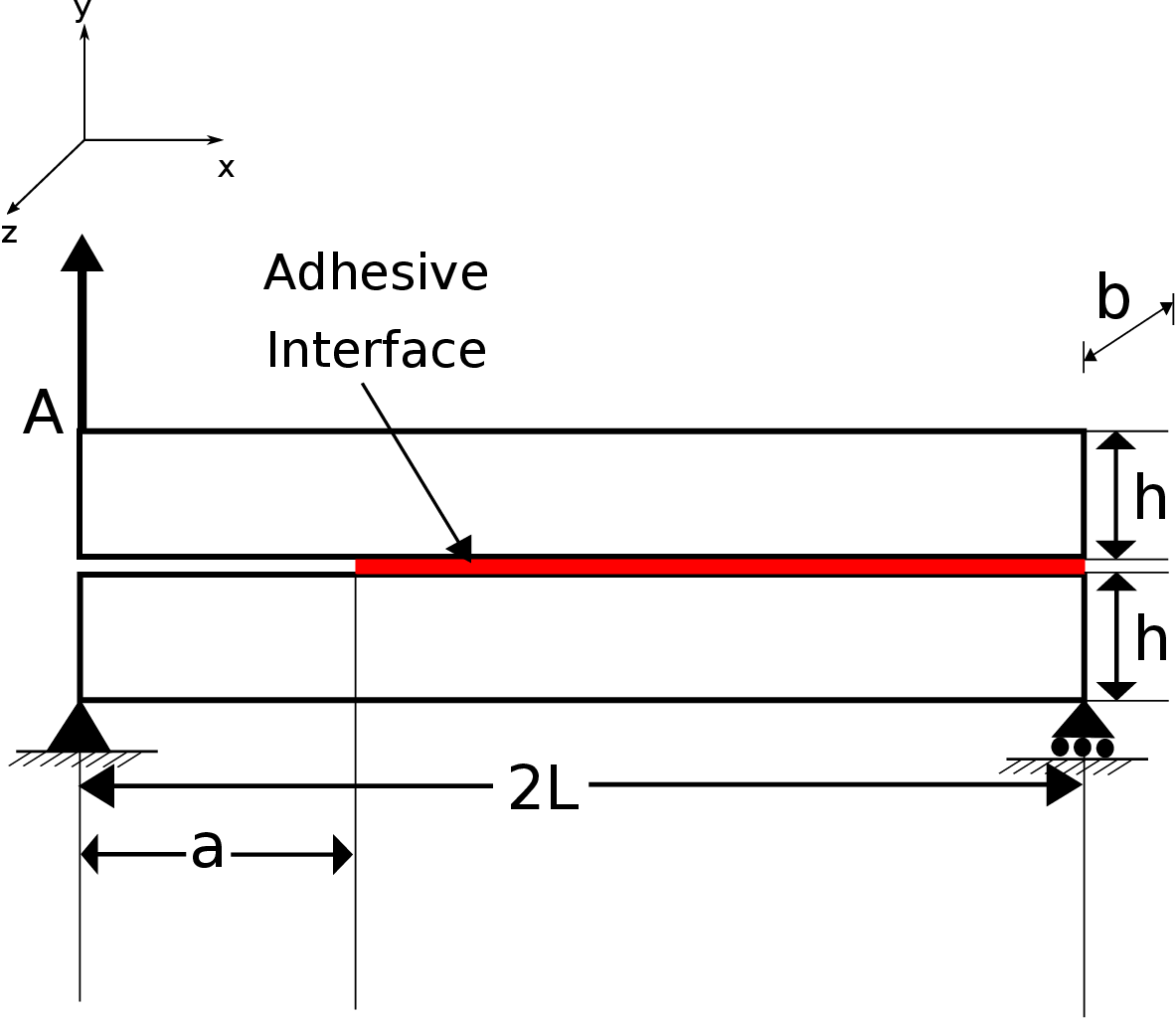}
    \caption{Setup of D.C.B specimen}
    \label{fig:f8}
\end{figure}
 \\Newton-Rapson technique is used with a displacement increment of 0.1mm in each step, with a convergence criterion of $10^{-4}$ in force norm.\\
 \begin{figure}[h!]
     \centering
 \includegraphics[scale = 0.25]{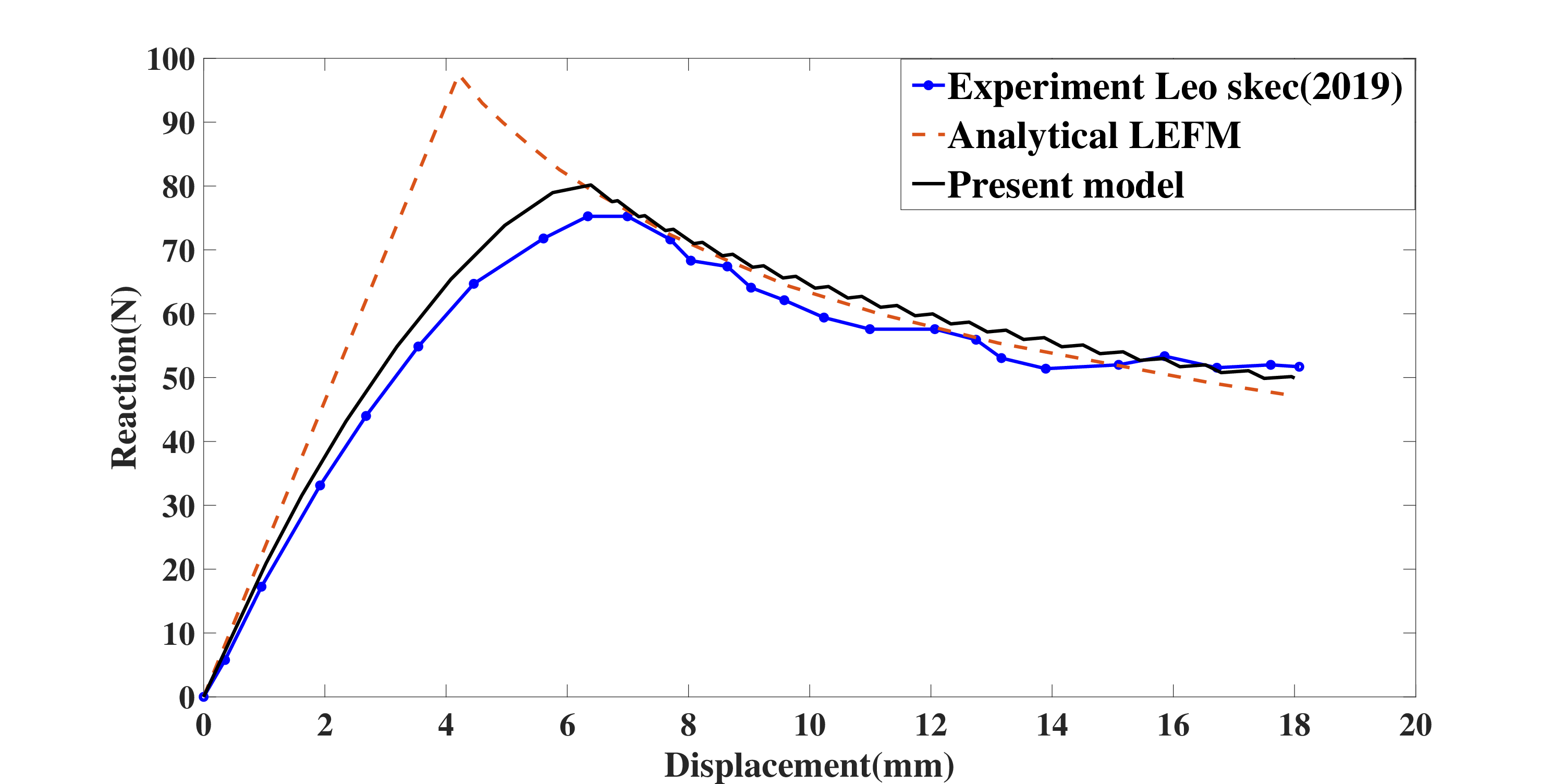}
     \caption{Comparison of simulation and experimental result~\cite{r31} of D.C.B specimen}
     \label{fig:f9}
 \end{figure}
 \\Figure \ref{fig:f9} shows the comparison of the experimental and numerical simulation results. It can be seen that the reaction displacement plot obtained by the proposed methodology is in close agreement with the experimental findings. The relative displacement of the interface contact nodes is calculated using the displacement and rotation of each node from the previously converged time step. Depending on the equations (6)-(14) the stiffness and energy release rate are chosen for the next load step.\\
 \\Since the displacement of any node in the X direction is negligible, for D.C.B simulation, stiffness in the mode-II direction is taken 100 times of mode-I stiffness, and $G_{IIc} > G_{Ic}$ is also taken.\\
 \begin{figure}[h!]
     \centering
 \includegraphics[scale = 0.25]{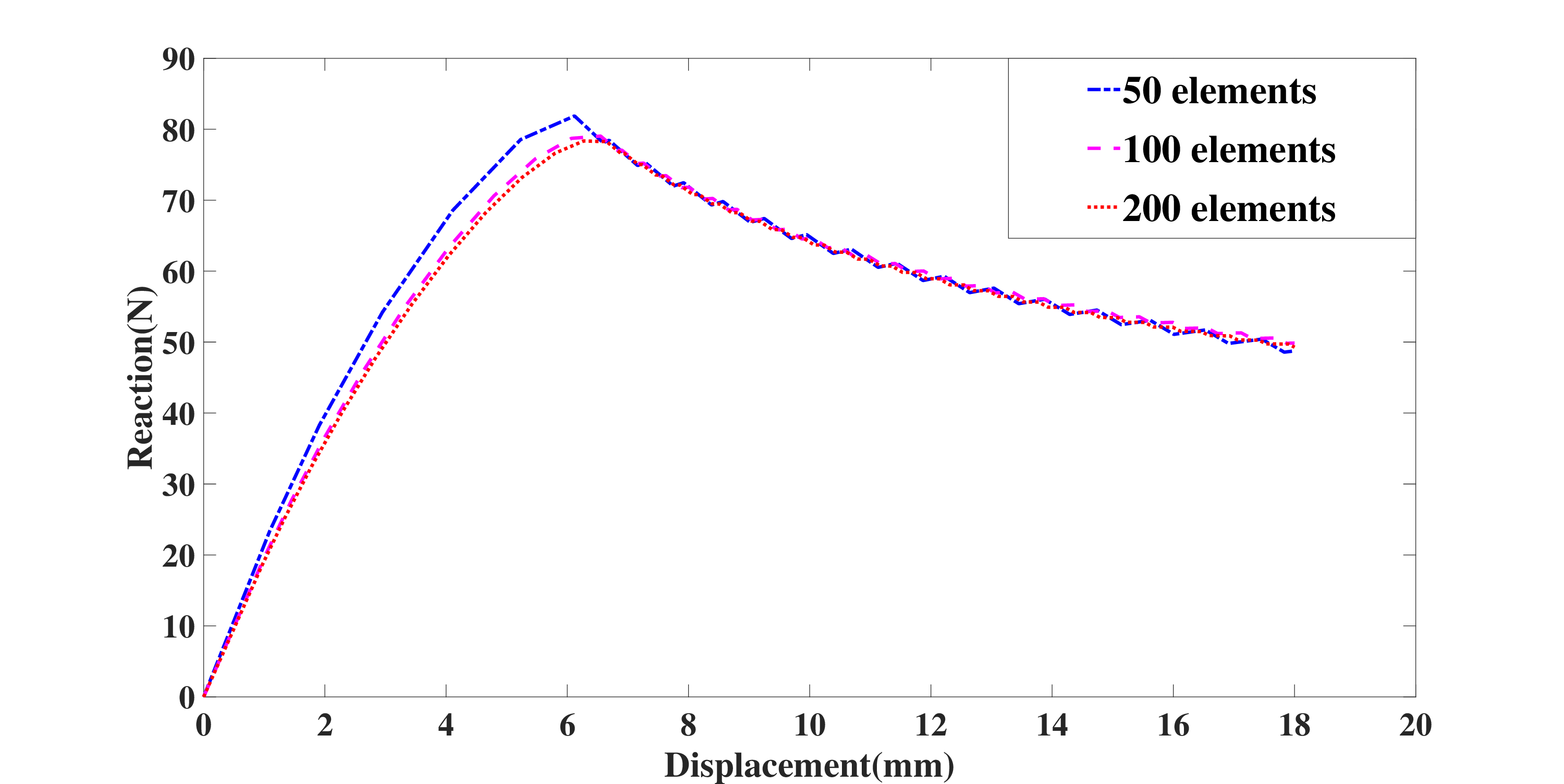}
     \caption{Mesh convergence study for D.C.B specimen}
     \label{fig:f10}
 \end{figure}
\\ Figure ~\ref{fig:f10} shows the mesh convergence study for the D.C.B specimen. it can be seen that the convergence is achieved with 100 elements on each beam. If the same problem is to be solved by the DCZM proposed by ~\cite{r17}, adherent would have meshed with a 2-D plane stress element. The degree of freedom will be huge for a  thin beam or laminate, but the proposed model overcomes this issue. As a result, the computational cost reduces from 8 unknowns in linear 2-D element to 6 unknowns in linear beam element.\\
\\Figure \ref{fig:f11} shows the end notch flexure (E.N.F) specimen setup, a three-point bending test with an initial crack length(a) at one end from the support. Since in three-point bending, the shear force is constant from support to the loading point, and shear stress is maximum at the middle of the beam. Hence, the crack will only advance due to shear stress, which ensures the Mode-II fracture. At point A, displacement is prescribed, and the reaction at the same location is evaluated.\\
\begin{figure}[h!]
    \centering
    \includegraphics[scale = 0.35]{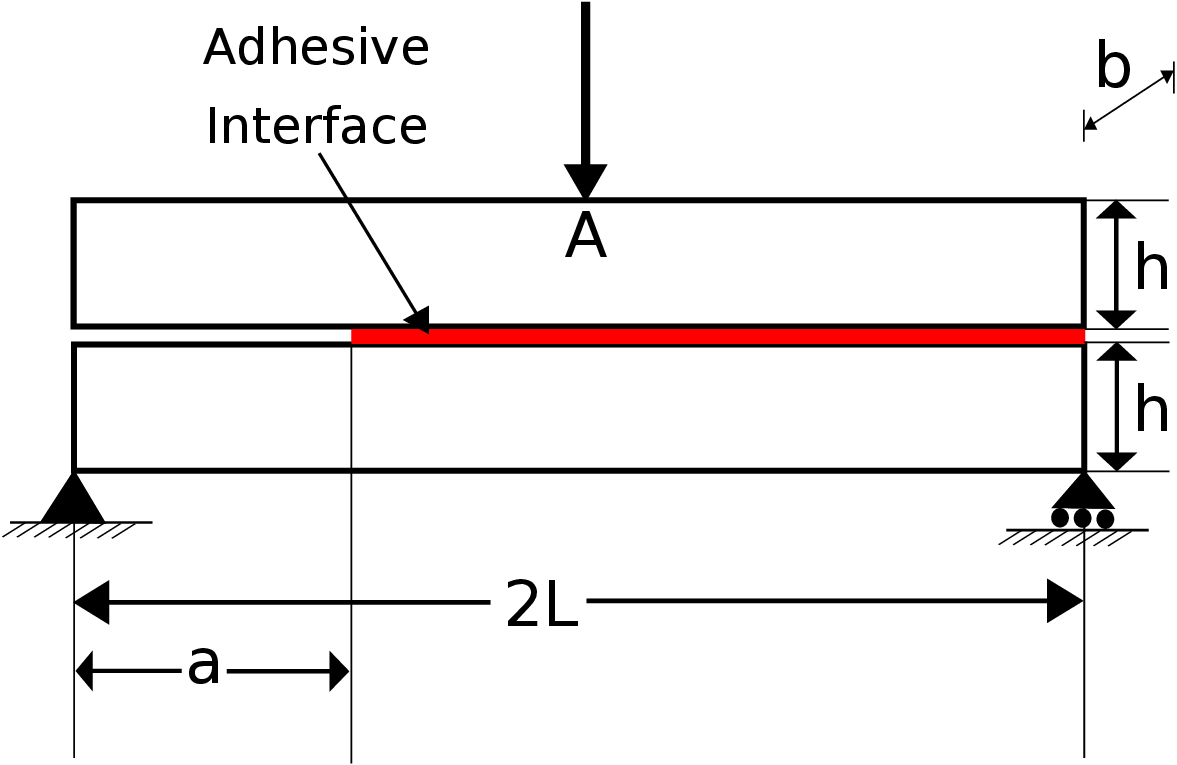}
    \caption{Setup of a E.N.F specimen}
    \label{fig:f11}
\end{figure}
\\A model with the same properties as a reference is chosen to validate the proposed scheme.The dimension of specimen is given as length (2L) = $100mm$, Width(b) = $1mm$, initial crack length(a) = $30mm$, height(h) =$1.5mm$, young’s modulus(E) = $135300N/mm^2$. The properties of adhesive is given as $G_{IIc} = 4N/mm$, $\tau_{c} = 57N/mm^2$, and $\gamma_c = 10^{-7}$.\\
\begin{figure}[h!]
    \centering
 \includegraphics[scale =0.25]{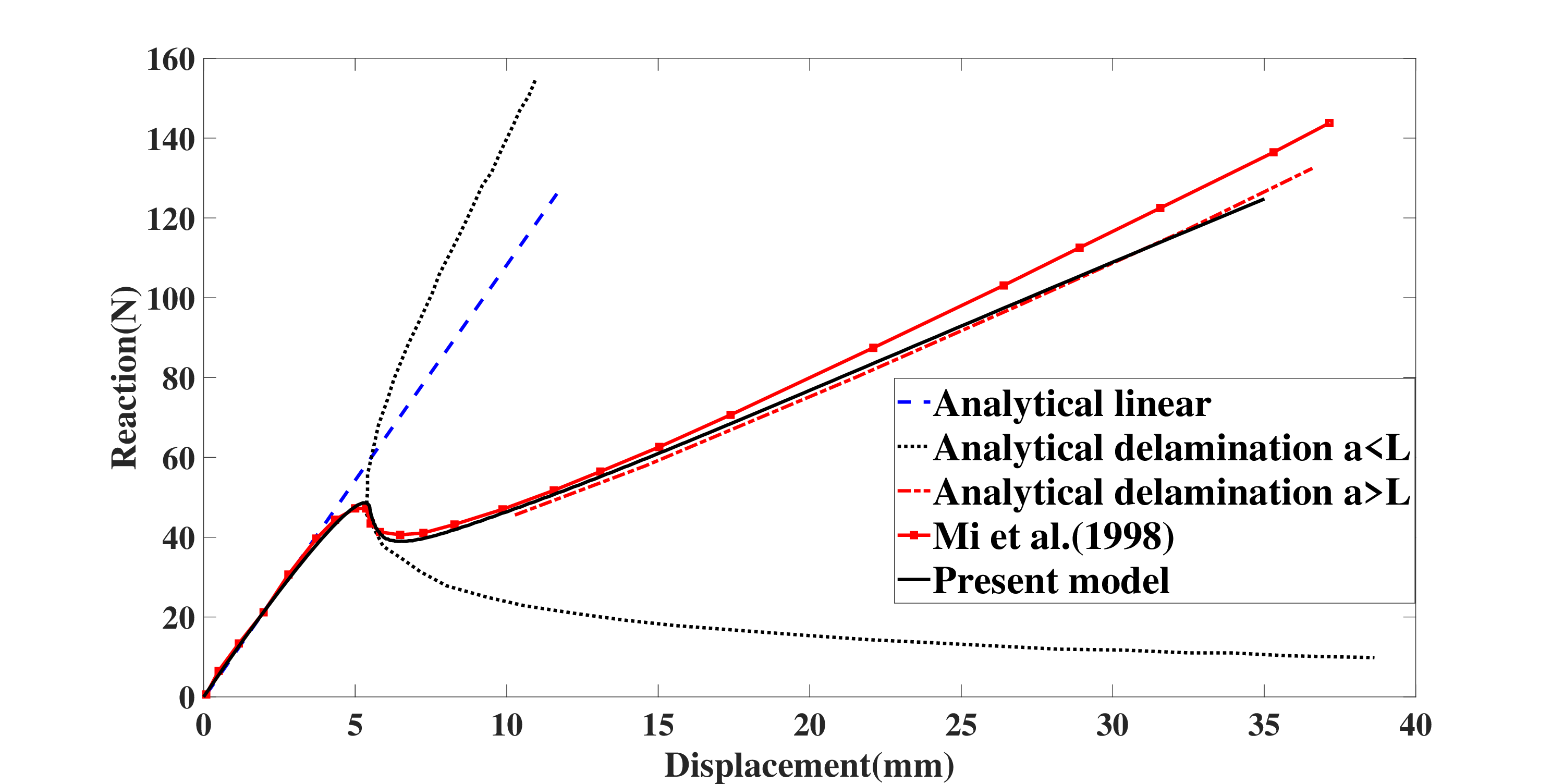}
    \caption{Reaction force for E.N.F specimen along with results by Y.Mi et al.~\cite{r10}}
    \label{fig:f12}
\end{figure}
\\Figure \ref{fig:f12} compares the numerical simulation of ~\cite{r10}. and the present work. Results are also in close agreement with the analytical results obtained from fracture mechanics. ~\cite{r10}. used 2-d elements to model the interface, which is computationally demanding and has a more stringent convergence requirement. The present work solves the same problem with the same accuracy, as shown in fig.6 with 606 unknown d.o.fs only. Current work also accurately captures bending behavior, as in a 2-d element, reduced integration is needed.\\
\begin{figure}[h!]
    \centering
\includegraphics[scale = 0.25]{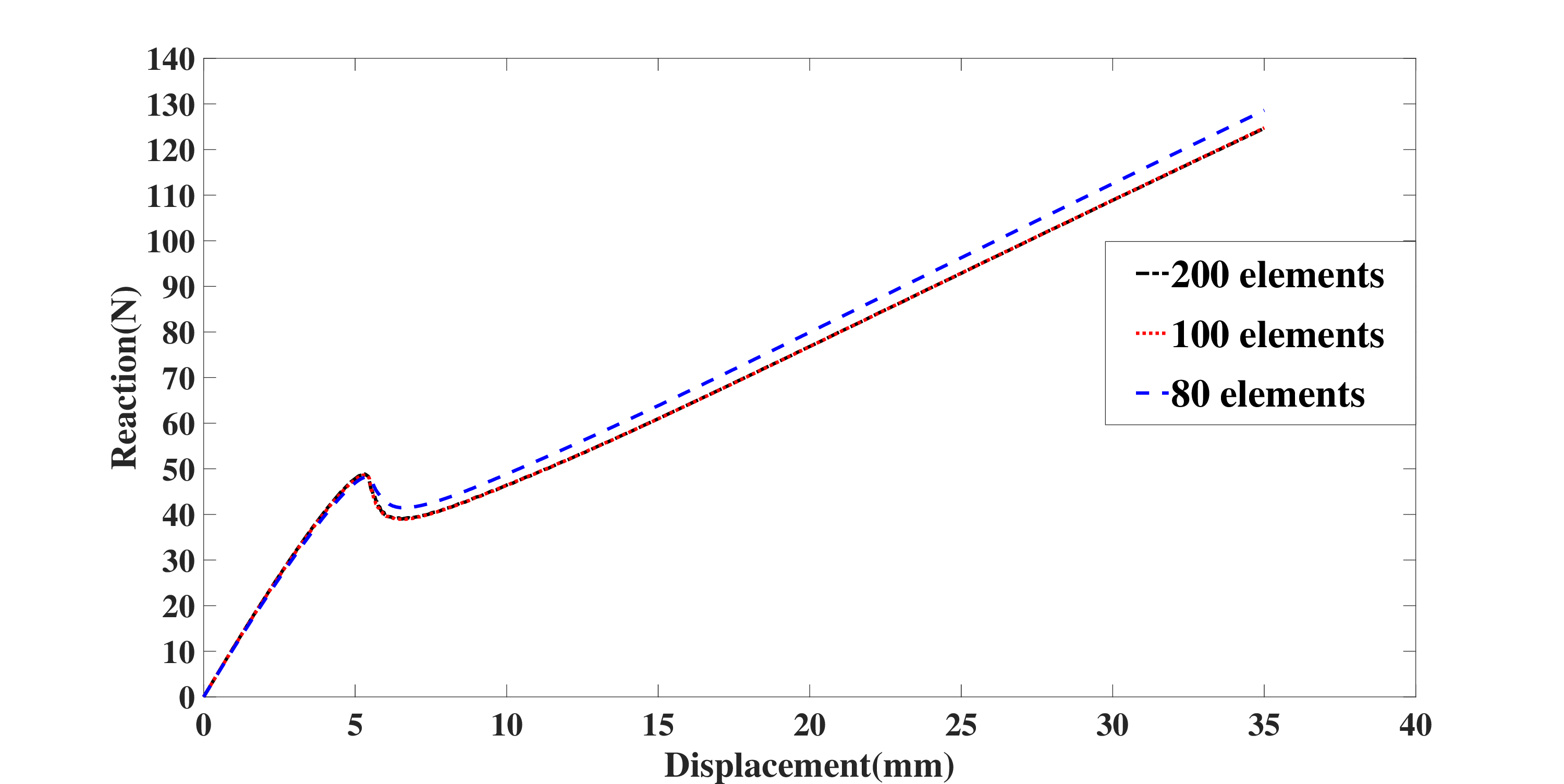}
    \caption{Mesh convergence study for E.N.F specimen}
    \label{fig:f13}
\end{figure}
\\The mesh convergence study for the E.N.F specimen is given in figure\ref{fig:f13}, and the Newton Rapson approach requires a minimum of 80 elements to achieve convergence. The results with 100 elements and 200 elements overlap each other, as seen in Figure \ref{fig:f13}. As a result, it is possible to conclude that 100 elements are adequate to solve this problem.\\
\begin{figure}[h!]
    \centering
\includegraphics[scale = 0.25]{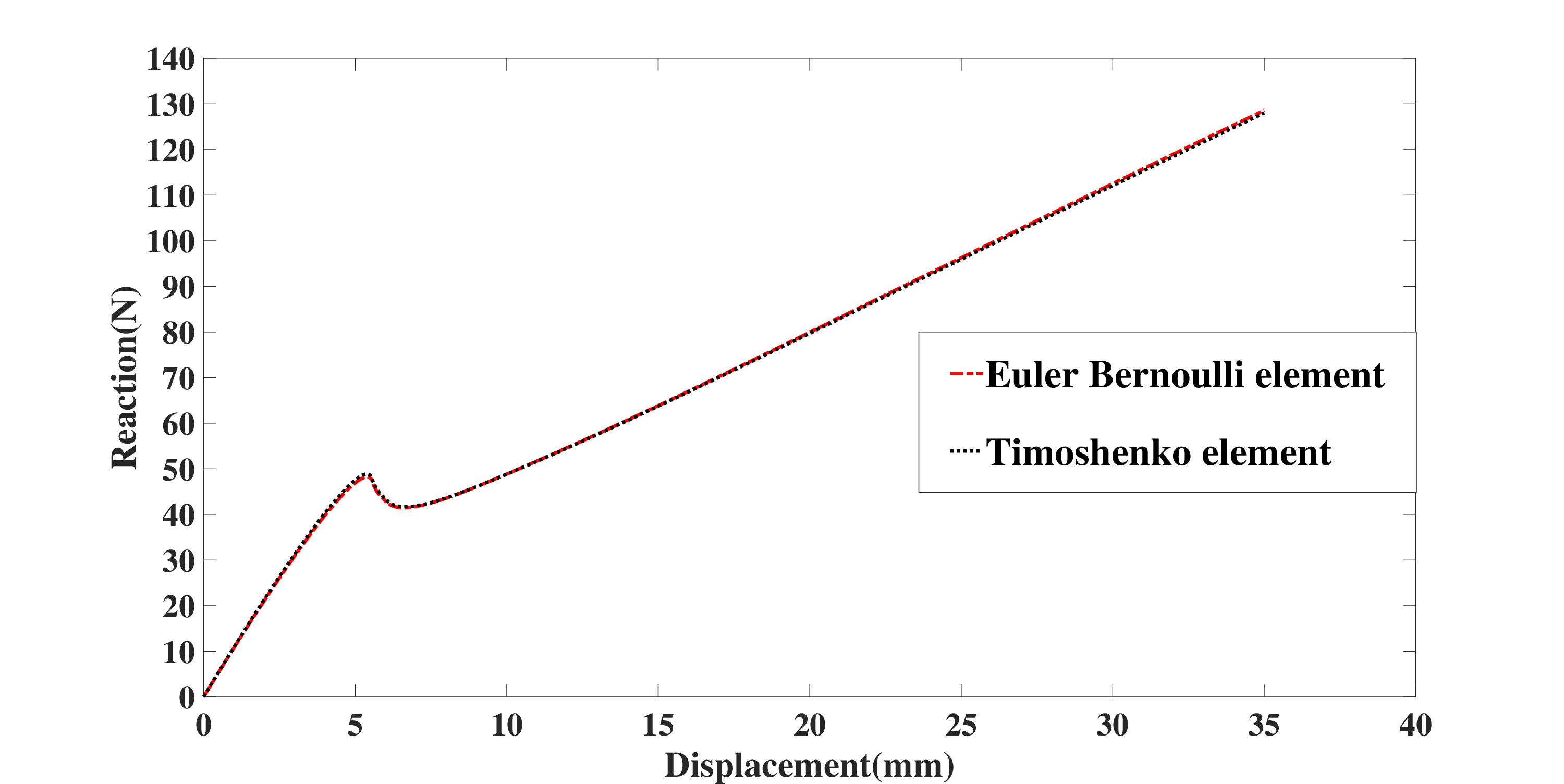}
    \caption{Comparison of Euler bernoulli and Timoshenko element}
    \label{fig:f14}
\end{figure}
\\Since Mode-II problems always involve shear stress, Figure \ref{fig:f14} compares simulations from the Timoshenko beam and Euler-Bernoulli beam elements. Figure \ref{fig:f14} shows both elements show an almost similar result, in case shear force is more dominant, Timoshenko beam can also be used.\\
\\For the mixed-mode analyses, the experimental setup of ~\cite{r32}, as shown in figure \ref{fig:f11}, has been simulated. By adjusting the load position on the lever($c$), a wide range of $G_{I}/G_{II}$ ratios can be achieved. The value of c is taken as 41.5mm to achieve $G_{I}/G_{II}$ = 1. While simulating the ratio of $G_{I}/G_{II}$ is related to the displacement boundary conditions given at points A and B, when this ratio equals unity, it means in each time increment, the displacement at both the points is increased by an equal amount but in the opposite direction. This ratio governs which mode is dominating while failure. \\
\begin{figure}[h!]
     \centering
     \includegraphics[scale = 0.25]{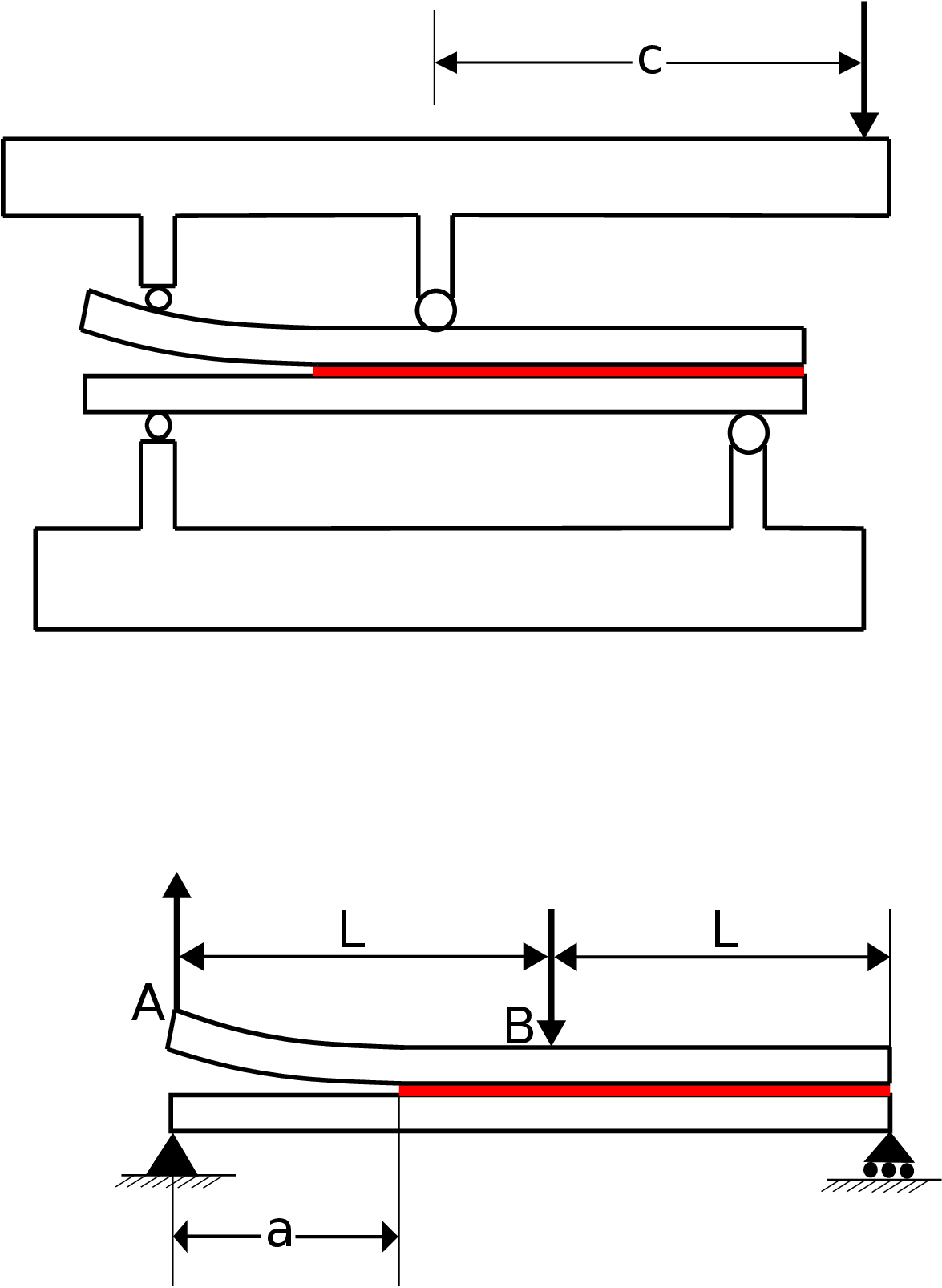}
     \caption{Setup of a M.M.B specimen}
     \label{fig:f15}
 \end{figure}
\\The model has following dimensions length (2L) = $100mm$, Width(b) = $1mm$, initial crack length(a) = $30mm$, height(h) =$1.5mm$, young’s modulus(E) = $135300N/mm^2$, and c = $41.5mm$. Properties of adhesive for both modes are taken the same as above D.C.B and E.N.F specimen.\\
\\Figure \ref{fig:f16} shows the reaction plot at point A against the applied displacement at the same point. As shown in figure \ref{fig:f16} the results have been validated with the analytical findings of ~\cite{r10}. Figure \ref{fig:f17} shows the mesh convergence study for the M.M.B specimen. It can be seen that even with 100 elements, satisfactory results are possible.\\
\begin{figure}[h!]
     \centering
 \includegraphics[scale = 0.25]{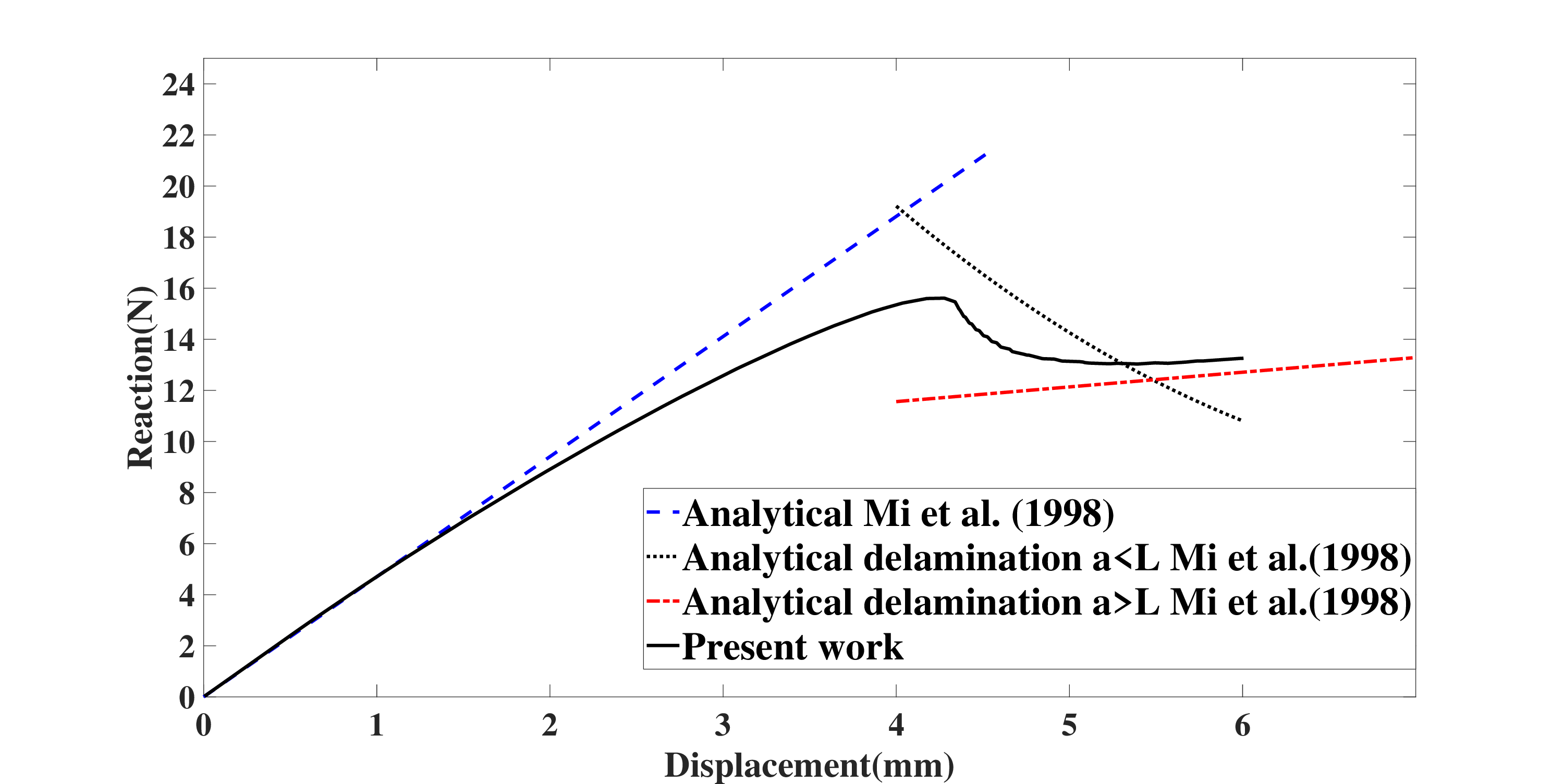}
     \caption{Reaction force for M.M.B specimen along with analytical results by ~\cite{r10}}
     \label{fig:f16}
 \end{figure}
 \begin{figure}[h!]
    \centering
 \includegraphics[scale = 0.25]{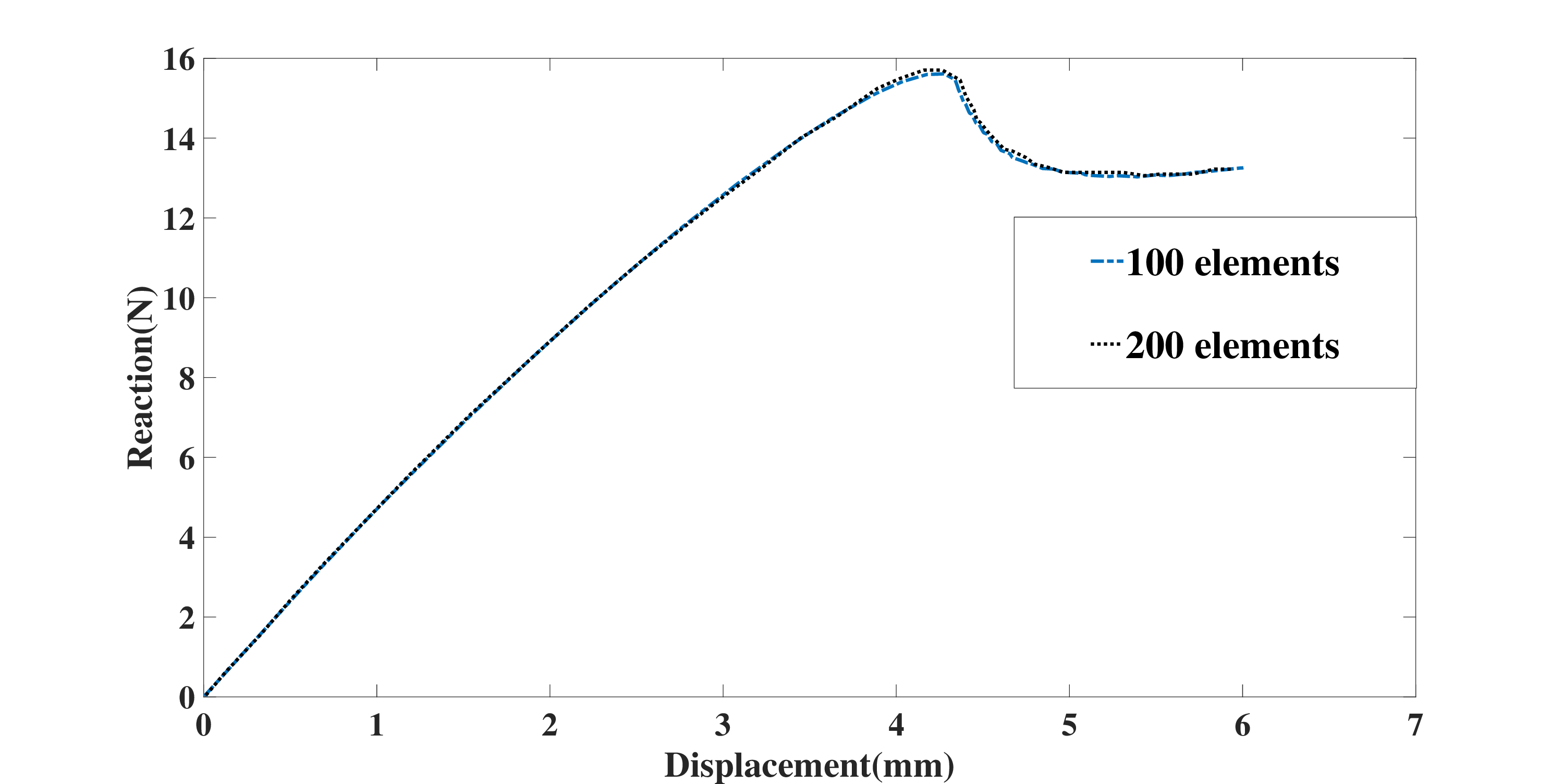}
    \caption{Mesh convergence plot for M.M.B simulations}
    \label{fig:f17}
\end{figure}
\\Two cases of interfacial crack propagation are simulated for a sandwich panel. In both simulations, the failure of the interface between the top skin and core is allowed. The first case replicates an experiment conducted by Prasad and Carlson (1994) using a Double Cantilever Beam (D.C.B) specimen subjected to Mode-I fracture. In the second case, an End-Notched Flexure (E.N.F) specimen is considered, following an experiment performed by Rinker et al. (2011), where failure is primarily influenced by a mode-II loading condition. The objective is to compare the accuracy of the proposed model with the experimental data available in the literature and evaluate its capability to describe the debonding failure mechanism in sandwich panels. The goal is to assess the model's performance and its ability to capture the behavior observed in the experimental studies.\\
\begin{figure}[h!]
     \centering
 \includegraphics[scale = 0.30]{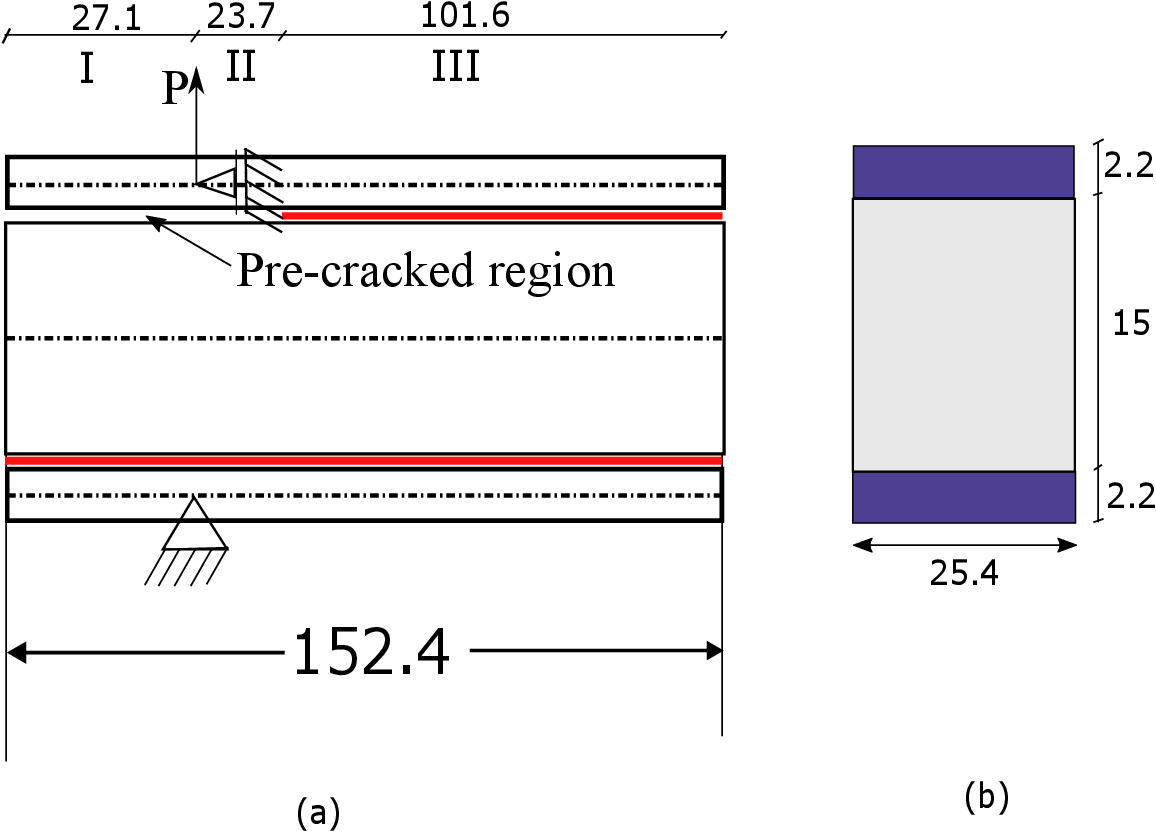}
     \caption{D.C.B setup of a sandwich beam (a) geometry (b) cross-section}
     \label{fig:f18}
 \end{figure}
\\Figure ~\ref{fig:f18} illustrates the configuration of a sandwich D.C.B panel. The top and bottom skins are composed of aluminum, while the core material is polymethacrylimide (type R90.400). In the modeling of this problem, the interface between the top skin and core is divided into three regions. The first two regions(I \& II as shown in figure ~\ref{fig:f18}) represent the pre-delaminated area, and they are separated by a support structure as depicted in figure ~\ref{fig:f18}. It is assumed that there is full bonding between the core and bottom skin along the entire length of the panel. In modeling, we consider that there is no contact between the top skin and core in the first and second regions. This assumption accounts for the absence of adhesive interaction or contact between these components in those specific areas. The Young's modulus and Poisson's ratio for the skin are considered as 70,000 MPa and 0.33, respectively. As for the core, the Young's modulus is taken as 420 MPa, and the Poisson's ratio is 0.25. The work of separation for both the interfaces is taken as $\varphi^{cT}$ = $\varphi^{cb} = 550$ $J/m^2$ and ${\delta_o^{cT}}$ = ${\delta_o^{cb}} = 0.12$ $mm$.\\
\begin{figure}[h!]
     \centering
 \includegraphics[scale = 0.25]{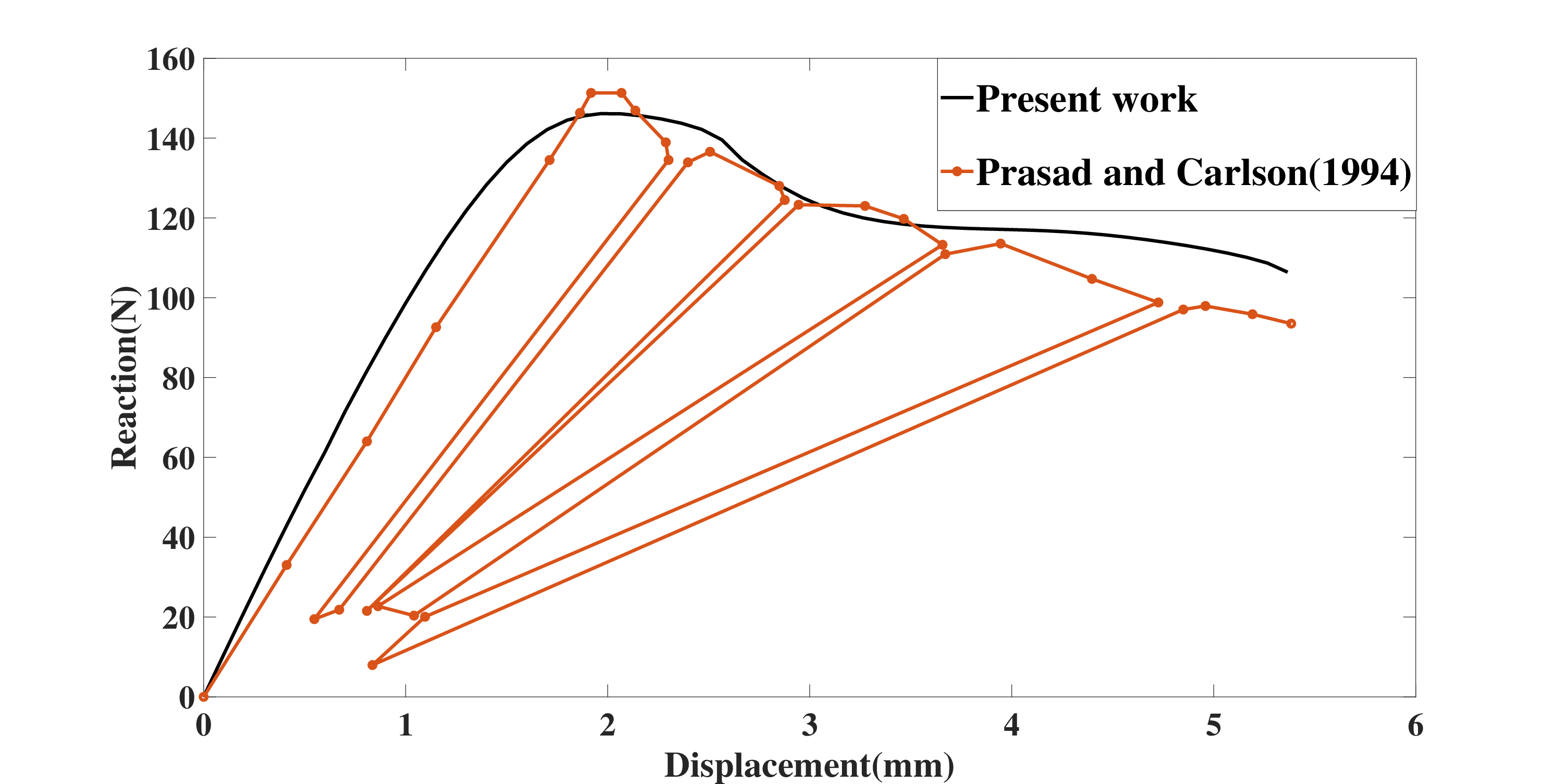}
     \caption{Comparison of simulation and experimental result of D.C.B sandwich specimen}
     \label{fig:f19}
 \end{figure}
\\ Figure ~\ref{fig:f19} illustrates the comparison between the experimental data and the results obtained from the present model. It is evident that the present model accurately predicts the behavior of the system, as the results closely align with the experimental findings.\\

\begin{figure}[h!]
     \centering
 \includegraphics[scale = 0.25]{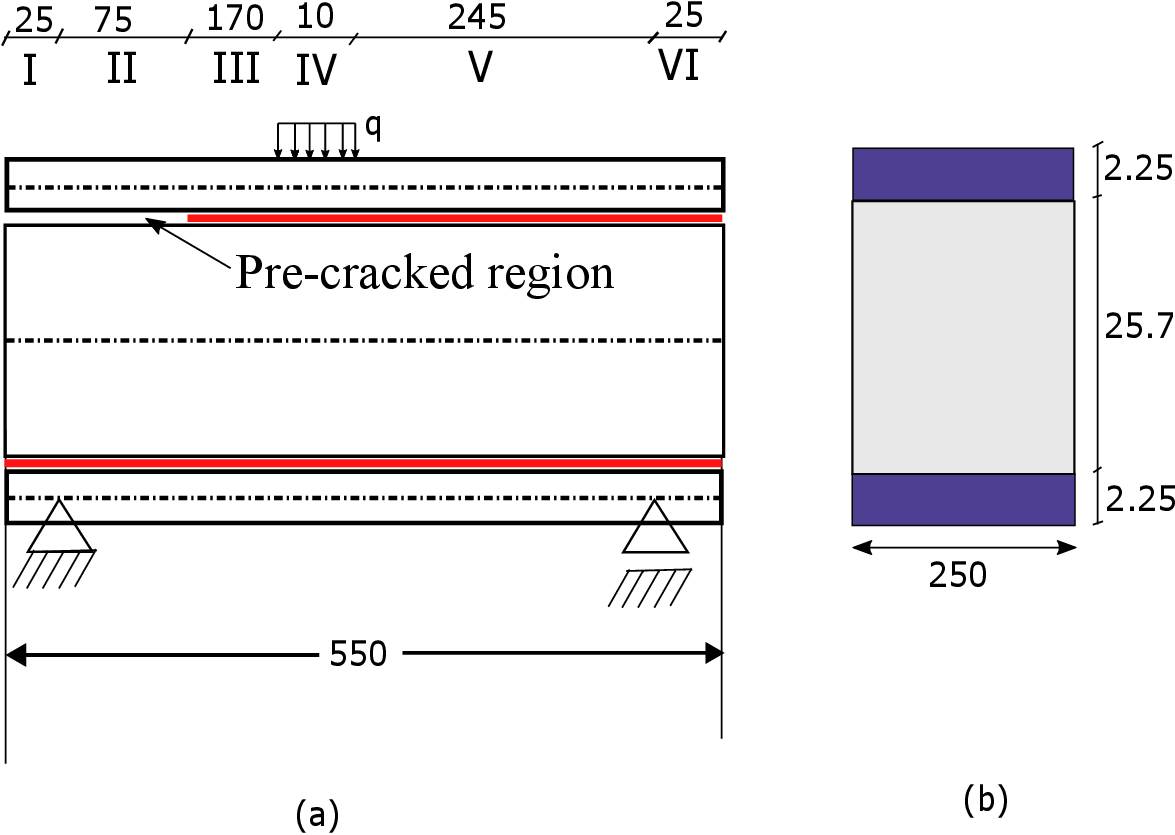}
     \caption{CSB setup of a sandwich beam (a) geometry (b) cross-section}
     \label{fig:f20}
 \end{figure}

Lastly, we conduct a simulation of an E.N.F sandwich specimen based on the experiment conducted by Rinker et al. (2011). Figure ~\ref{fig:f20} depicts the setup and geometry of the specimen, which is subjected to a three-point bending loading configuration. The central applied load is distributed over a length of 10 mm. The pre-cracked interface is allowed to propagate due to shear forces, resulting in a mode-II failure of the interface. The sandwich panel is divided into six zones, with the first and second zones separated by a support, representing the pre-delaminated region of the specimen. In this problem, contact between the pre-delaminated region and the underlying core occurs when the load is applied. To prevent interpenetration between the top skin and the core, the contact condition described by equation (18) is utilized.\\
\\The top and bottom skin is made up of CFRP composite laminates of arrangement $[(+45, 0, -45)_3]_s$. The material properties of laminates is given as $E_{xx} = 13500$ $MPa$, $E_{zz} = 9750$ $MPa$, $\nu_{xz} = 0.28$. The core is made up of Core is taken as $E^c = 105$ $MPa$, $G^c = 42$ $MPa$ and $\nu^c = 0.25$.The interface properties for top and bottom interface is also same and taken as $\varphi^{cT}$ = $\varphi^{cb} = 385$ $J/m^2$ and ${\delta_o^{cT}}$ = ${\delta_o^{cb}} = 0.2$ $mm$.\\
\begin{figure}[h!]
     \centering
 \includegraphics[scale = 0.25]{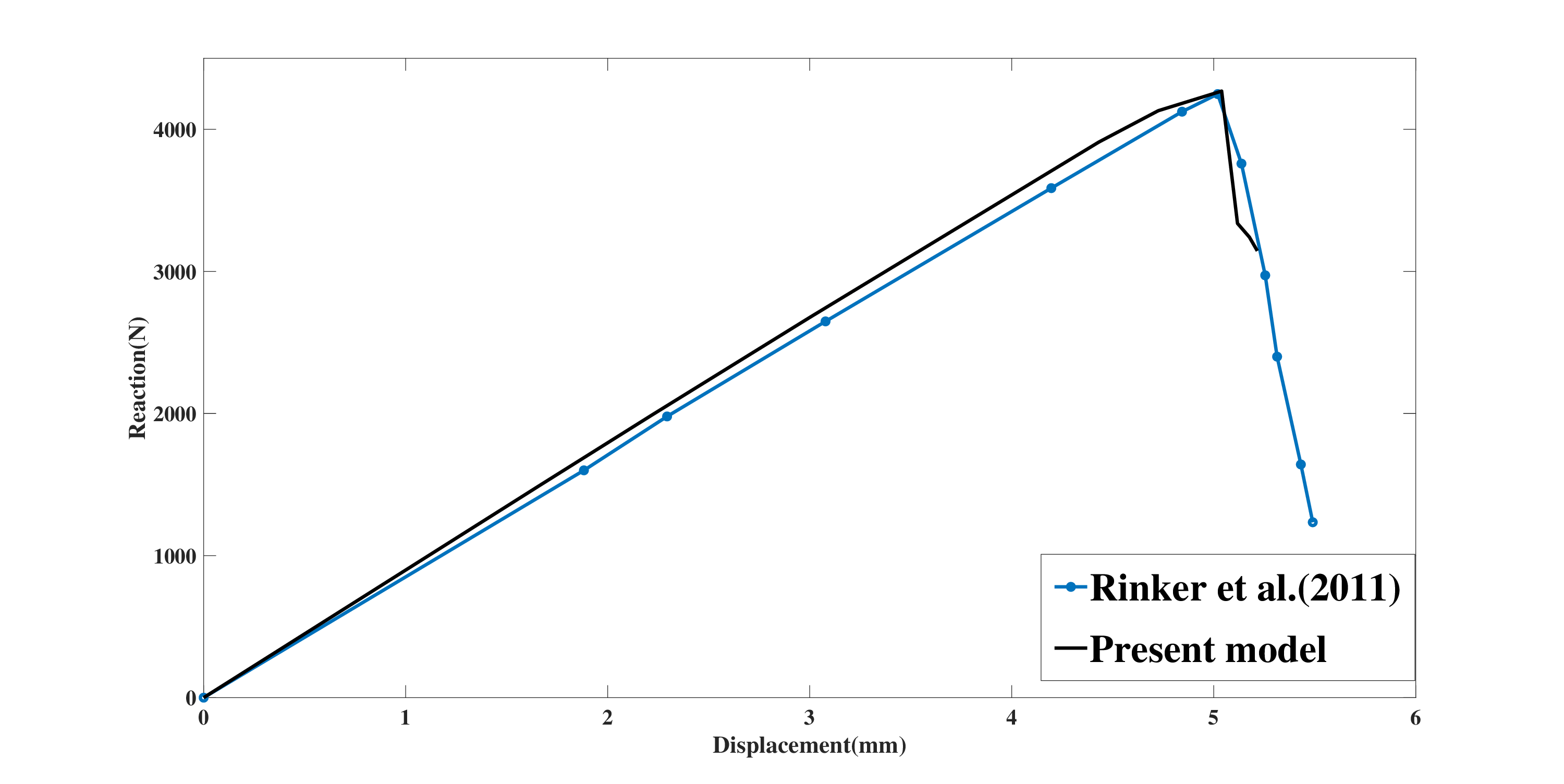}
     \caption{Comparison of simulation and experimental result of CSB specimen}
     \label{fig:f21}
 \end{figure}
\\Figure ~\ref{fig:f21} shows the present model predict very close result with the experimental data available in the literature. 
%$
%\\
%\\
% Numbered list
% Use the style of numbering in square brackets.
% If nothing is used, default style will be taken.
%\begin{enumerate}[a)]
%\item 
%\item 
%\item 
%\end{enumerate}  

% Unnumbered list
%\begin{itemize}
%\item 
%\item 
%\item 
%\end{itemize}  

% Description list
%\begin{description}
%\item[]
%\item[] 
%\item[] 
%\end{description}  

% Figure
%\begin{figure}[<options>]
%	\centering
		%\includegraphics[<options>]{}
	%  \caption{}\label{fig1}
%\end{figure}

%\begin{table}[<options>]
%\caption{}\label{tbl1}
%\begin{tabular*}{\tblwidth}{@{}LL@{}}
%\toprule
%  &  \\ % Table header row
%\midrule
% & \\
% & \\
% & \\
% & \\
%\bottomrule
%\end{tabular*}
%\end{table}

% Uncomment and use as the case may be
%\begin{theorem} 
%\end{theorem}

% Uncomment and use as the case may be
%\begin{lemma} 
%\end{lemma}

%% The Appendices part is started with the command \appendix;
%% appendix sections are then done as normal sections
%% \appendix
\section{Summary and Conclusion}
A novel beam-based discrete cohesive zone model has been developed and its capabilities in simulating the initiation and propagation of cracks has been evaluated. The proposed methodology has been validated through benchmark simulations, including the Double Cantilever Beam (D.C.B), End Notched Flexure beam (E.N.F), and Mixed Mode Bending (M.M.B) tests. The obtained results have been compared with the corresponding findings reported in the literature, confirming the accuracy and reliability of the model. In contrast to the cohesive zone models previously reported in the literature, which treat adherents as continuum elements in both 2D and 3D configurations, the present approach addresses the concern of mesh quality and refinement, particularly for thin laminates. By employing the current scheme, computationally expensive techniques for thin laminates are circumvented. Moreover, the current formulation facilitates the accurate representation of bending behavior, which can be challenging to capture using 2D continuum elements.\\

Furthermore, the model has been extended to encompass a three-layer sandwich panel, consisting of a top skin, core, and bottom skin. Considering that the core undergoes shear deformation while the skins primarily experiences bending forces due to their distant location from the neutral axis, a higher order beam element is utilized to model the core, whereas the Timoshenko element is employed for the skins. The failure at the interface between the core and skins is thoroughly investigated, and the obtained results are validated against experimental data from the literature.

Thus, this work presents a beam-based discrete cohesive zone model that offers a comprehensive analysis of the interface failure in both thin laminates and sandwich panels with computational efficacy and accuracy.
\section*{Appendix}
Consider a rectangular cross section beam of length (L) , width (b) and height ($2h^{c}$). The axial and transverse displacement of beam is approximated using Taylor's series expansion. For axial displacement first four terms of Taylor's expansion  are considered whereas for transverse displacement first three terms are considered.  
\begin{figure}[htp]
    \centering
    \includegraphics[scale = 0.5]{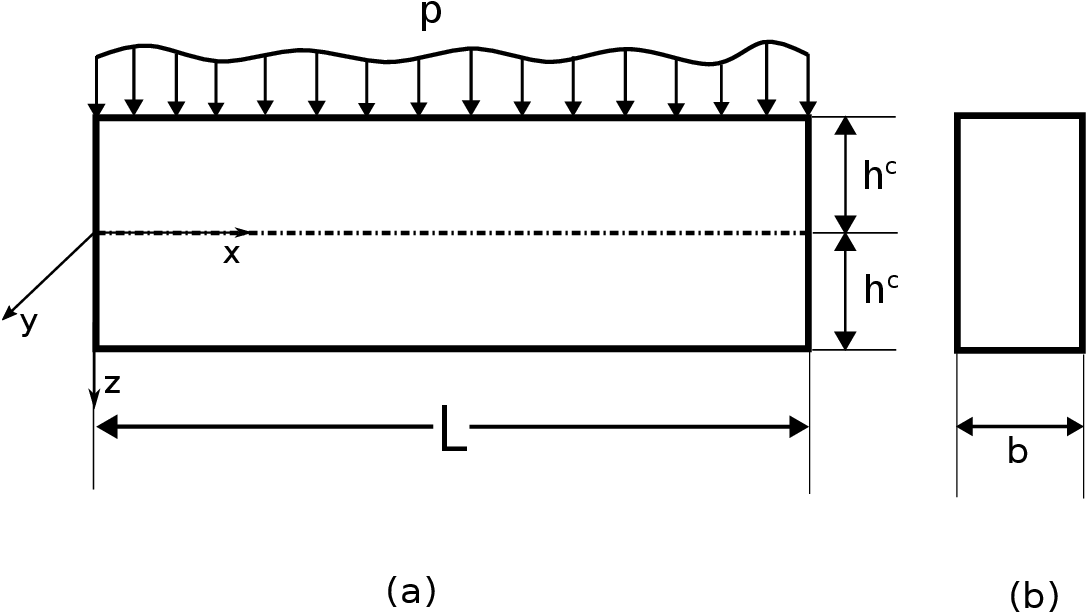}
    \caption{(a) geometric representation of beam with applied load (b) cross section of beam}
    \label{fig:galaxy}
\end{figure}

\begin{equation}
 U(x,z^c) = u_o^c(x) + z^cu_1^c(x) + (z^c)^2u_2^c(x) + (z^c)^3u_3^c(x) \tag{A.1}
 \end{equation}
 \begin{equation}
 W(x,z^c) = w_o^c(x) + z^cw_1^c(x) + (z^c)^2w_2^c(x) \tag{A.2}
 \end{equation}
Displacement along width direction is taken as 0. Upon using linear strain displacement relationship the components of strain tensor can be evaluated and are given as 
\begin{equation}
\epsilon_{xx} = \frac{\partial U(x,z^c)}{\partial x} = u_o^c,_x + z^cu_1^c,_x + (z^c)^2u_2^c,_x + (z^c)^3u_3^c,_x \tag{A.3}
\end{equation}
\begin{equation}
\epsilon_{zz} = \frac{\partial W(x,z^c)}{\partial z^c} =   w_1^c + (2z^c)w_2^c \tag{A.4}
\end{equation}
\begin{equation}
\gamma_{xz} = \frac{\partial U(x,z^c)}{\partial z^c} + \frac{\partial W(x,z^c)}{\partial x} = (u_1^c + w_o^c,_x) + z^c(2u_2^c + w_1^c,_x) + (z^c)^2(3u_3^c + w_2^c,_x) \tag{A.5} 
\end{equation}
\begin{equation}
\gamma_{xy} = \gamma_{yz} = \epsilon_{yy} = 0 \tag{A.6}
\end{equation}
 Assuming the beam is made up of isotropic linear elastic material so the constitutive relationship can be given as:
 \begin{equation}
\begin{Bmatrix}
\sigma_{xx} \\
\sigma_{zz} \\
\tau_{xz} \\
\end{Bmatrix}
=
\begin{bmatrix}
A_{11} & A_{12} & 0 \\
A_{21} & A_{22} & 0 \\
0 & 0 & A_{33} \\
\end{bmatrix}
\begin{Bmatrix}
\epsilon_{xx} \\
\epsilon_{zz} \\
\gamma_{xz} \\
\end{Bmatrix}   \tag{A.7}
\end{equation}
For this beam the potential energy is given as
\begin{equation}
\Pi = \mathbb{U - V}\\ \tag{A.8}
\end{equation}
$\mathbb{U}$: internal strain energy\\
$\mathbb{V}$: external work done\\
and from principle of stationary potential energy 
\begin{equation}
\delta\Pi = \mathbb{\delta  U} - \mathbb{\delta  V} = 0 \tag{A.9}
\end{equation}
\begin{equation}
\mathbb{\delta U} =  \int_{V} (\sigma_{xx}\delta\epsilon_{xx} + \sigma_{zz}\delta\epsilon_{zz} + \tau_{xz}\delta\gamma_{xz}) dV \tag{A.10}
\end{equation} 
Upon using (A.3) - (A.6), the each term of (A.10) can be expanded as 
\begin{equation}
\int_{V} \sigma_{xx}\delta\epsilon_{xx} dV = \int_{0}^{L} N_{xx}\delta u_{o}^c,_{x} dx + \int_{0}^{L} M_{xx}\delta u_{1}^c,_{x} dx\\
+ \int_{0}^{L} M_{xx1}\delta u_{2}^c,_{x} dx + \int_{0}^{L} M_{xx2}\delta u_{2}^c,_{x} dx \tag{A.11}
\end{equation}
\begin{equation}
\int_{V} \sigma_{zz}\delta\epsilon_{zz} dV = \int_{0}^{L} R_{zz}\delta w_{1}^c dx + 2\int_{0}^{L} M_{zz}\delta w_{2}^c dx \tag{A.12}
\end{equation}
\begin{equation}
\int_{V} \tau_{xz}\delta\gamma_{xz} dV = \int_{0}^{L} V_{xz}(\delta u_{1}^c + \delta w_o^c,_{x}) + \int_{0}^{L} V_{xz1}(2\delta u_{2}^c + \delta w_1^c,_{x}) + \int_{0}^{L} V_{xz2}(3\delta u_{3}^c + \delta w_2^c,_{x}) \tag{A.13}
\end{equation}
where these stress resultant can be calculated upon performing integral across the cross section as 
\begin{equation}
N_{xx} = \int_{A} \sigma_{xx} dA \tag{A.14}
\end{equation}
\begin{equation}
    M_{xx} = \int_{A} z^c\sigma_{xx} dA \tag{A.15}
\end{equation}
\begin{equation}
    M_{xx1} = \int_{A} (z^c)^2\sigma_{xx} dA \tag{A.16}
\end{equation}
\begin{equation}
    M_{xx2} = \int_{A} (z^c)^3\sigma_{xx} dA \tag{A.17}
\end{equation}
\begin{equation}
    R_{zz} = \int_{A} \sigma_{zz} dA \tag{A.18}
\end{equation}
\begin{equation}
    M_{zz} = \int_{A} z^c\sigma_{zz} dA \tag{A.19}
\end{equation}
\begin{equation}
    V_{xz} = \int_{A} \tau_{xz} dA \tag{A.20}
\end{equation}
\begin{equation}
    V_{xz1} = \int_{A}z^c \tau_{xz} dA \tag{A.21}
\end{equation}
\begin{equation}
    V_{xz2} = \int_{A}(z^c)^2 \tau_{xz} dA \tag{A.22}
\end{equation}
upon using (A.7), (A.3) - (A.6) and (A.14) - (A.22) the stress resultants can be represented as 
\begin{equation}
\bar{\sigma} = D\bar{\epsilon} \tag{A.23}
\end{equation}
where,
\begin{equation}
\bar{\sigma} = 
\begin{Bmatrix}
N_{xx}\quad
M_{xx}\quad
M_{xx1}\quad
M_{xx2}\quad
R_{zz}\quad
2M_{zz}\quad
V_{xz}\quad
V_{xz1}\quad
V_{xz2}
\end{Bmatrix}^t \tag{A.24}
\end{equation}
\begin{equation}
\bar{\epsilon} = 
\begin{Bmatrix}
u_{o}^c,_{x}\quad
u_{1}^c,_{x}\quad
u_{2}^c,_{x}\quad
u_{3}^c,_{x}\quad
w_{1}^c\quad
w_{2}^c\quad
u_{1}^c + w_{o}^c,_{x}\quad
2u_{2}^c + w_{1}^c,_{x}\quad
3u_{3}^c + w_{2}^c,_{x}
\end{Bmatrix} ^t  \tag{A.25}
\end{equation}
\begin{equation}
D = 
\begin{bmatrix}
    A_{11}A & 0 & A_{11}I & 0 & A_{12}A & 0 & 0 & 0 & 0 \\
    0 & A_{11}I & 0 & A_{11}I_{1} & 0 & 2A_{12}I & 0 & 0 & 0 \\
    A_{11}I & 0 & A_{11}I_{1} & 0 & A_{12}A & 0 & 0 & 0 & 0 \\
    0 & A_{11}I_{1} & 0 & A_{11}I_{2} & 0 & 2A_{12}I_{1} & 0 & 0 & 0 \\
    A_{21}A & 0 & A_{21}I & 0 & A_{33}A & 0 & 0 & 0 & 0 \\
    0 & 2A_{21}I & 0 & 2A_{21}I_{1} & 0 & 4A_{33} & 0 & 0 & 0 \\
    0 & 0 & 0 & 0 & 0 & 0 & A_{33}A & 0 & A_{33}I \\
    0 & 0 & 0 & 0 & 0 & 0 & 0 & A_{66}I & 0 \\
    0 & 0 & 0 & 0 & 0 & 0 & A_{33}A & 0 & A_{33}I_{1} \\
\end{bmatrix} \tag{A.26}
\end{equation}
where 
\begin{equation}\tag{A.27}
\begin{split}
 A = b(2h^c)\quad \quad I = b\int_{-h^c}^{h^c} (z^c)^2 dz^c\quad \quad  I_{1} = b\int_{-h^c}^{h^c} (z^c)^4 dz^c\quad \quad I_{2} = \int_{-h^c}^{h^c} (z^c)^6 dz^c 
\end{split}
\end{equation}
Hence, the internal strain energy can be represented as
\begin{equation}\tag{A.28}
\delta\mathbb{U} = \int_{0}^{L} \delta\bar{\epsilon}^t\bar{\sigma} dx = \int_{0}^{L} \delta\bar{\epsilon}^tD\bar{\epsilon} dx
\end{equation}
As shown in fig $p$ donates the force per unit length acting on the upper face of the beam i.e., at $z^c$ = $-h^c$. $p$ contains two component of force $p_z(x)$ and $p_u(x)$ one in  transverse and other in axial direction. Therefore the work done by external forces can be given as 
\begin{equation} \tag{A.29}
    \delta\mathbb{V} = \int_{0}^{L}p^t\delta\hat{u}^c dx
\end{equation}
where,
\begin{equation}\tag{A.30}
    p = \begin{Bmatrix}
        p_u\quad -p_uh^c\quad p_u(h^c)^2\quad -p_u(h^c)^3\quad p_w\quad -p_wh^c\quad p_w(h^c)^2
    \end{Bmatrix}^t
\end{equation}
\begin{equation} \tag{A.31}
    \delta\hat{u}^c = \delta\begin{Bmatrix}
        u_{o}^c\quad u_{1}^c\quad u_{2}^c\quad u_{3}^c\quad w_{o}^c\quad w_{1}^c\quad w_{2}^c
    \end{Bmatrix}^t
\end{equation}
\subsection*{Finite element formulation}
Let the beam is discretized into n number of 2 noded linear elements. Then the interpolation of displacement in $e^{th}$ element from the neighbouring nodal displacement of $e^{th}$ and $(e+1)^{th}$ node is given as
\begin{equation} \tag{A.32}
    \hat{u}^c = \begin{bmatrix}
                  N
                  \end{bmatrix}_{7*14}
               \begin{Bmatrix}
                 \hat{U}^{ce}
               \end{Bmatrix}_{14*1}
\end{equation}
\begin{equation}\tag{A.33}
\begin{bmatrix}
N
\end{bmatrix} = \begin{bmatrix}
                  N_1[\mathbb{I}]_{7*7}\quad N_2[\mathbb{I}]_{7*7}
                \end{bmatrix}
\end{equation}
\begin{equation} \tag{A.34}
\begin{Bmatrix}
\hat{U}^{ce}
\end{Bmatrix} = \begin{Bmatrix}
                \{\hat{u}^{c(e)}\}_{7*1}\\
               \{\hat{u}^{c(e+1)}\}_{7*1}\\       
                \end{Bmatrix}
\end{equation}
where $N_1$ and $N_2$ are the Lagrange linear interpolation functions. ${u}^{c(e)}$ and ${u}^{c(e+1)}$ are the values of $\hat{u}^{c}$ at $e^{th}$ and $(e+1)^{th}$ nodes and $\mathbb{I}$ is the identity matrix.\\
Further the strain resultants can also be represented in terms of displacements as
\begin{equation} \tag{A.35}
\bar{\epsilon} = L\hat{u}^c = B\hat{U}^{ce} 
\end{equation}
where, $B=LN$
\begin{equation} \tag{A.36}
    L = \begin{bmatrix}
        \frac{d}{dx} & 0 & 0 & 0 & 0 & 0 & 0 \\
        0 & \frac{d}{dx} & 0 & 0 & 0 & 0 & 0 \\
        0 & 0 & \frac{d}{dx} & 0 & 0 & 0 & 0 \\
        0 & 0 & 0 & \frac{d}{dx} & 0 & 0 & 0 \\
        0 & 0 & 0 & 0 & 0 & 1 & 0 \\
        0 & 0 & 0 & 0 & 0 & 0 & 1 \\
        0 & 1 & 0 & 0 & \frac{d}{dx} & 0 & 0 \\
        0 & 0 & 2 & 0 & 0 & \frac{d}{dx} & 0 \\
        0 & 0 & 0 & 3 & 0 & 0 & \frac{d}{dx} \\
       \end{bmatrix}_{9*7}
\end{equation}
hence for one element the internal strain energy is given by
\begin{equation} \tag{A.37}
    \delta{\mathbb{U}} = \int_{L_e}^{} (\delta\hat{U}^{ce})^t[B^tDB]\begin{Bmatrix} \hat{U}^{ce} \end{Bmatrix} dx
\end{equation}
the work done by external force on this particular element is given as
\begin{equation} \tag{A.38}
    \delta{\mathbb{V}} = \int_{L_e}^{} \delta\hat{U}^{ce}N^{t}p dx
\end{equation}
Hence for arbitrary $\delta\hat{U}^{ce}$ from variational principle 
\begin{equation} \tag{A.39}
    K^c\hat{U}^{ce} = F^c
\end{equation}
the element stiffness matrix and external force vector is given as 
    \begin{equation} \tag{A.40}
    K^c = \int_{L_e}^{}B^tDB dx 
    \end{equation}
    \begin{equation} \tag{A.41}
    F^c = \int_{L_e}^{} N^tp dx 
\end{equation}

% To print the credit authorship contribution details
%\printcredits

%% Loading bibliography style file
%\bibliographystyle{model1-num-names}
% \bibliographystyle{cas-model1-num-names}

% Loading bibliography database
% \bibliography{ijf_bib}

% Biography
%\bio{}
% Here goes the biography details.
% \endbio

%\bio{pic1}
% Here goes the biography details.
% \endbio

% % end of file template.tex

\end{document}